\documentclass[twocolumn]{aastex63}

\usepackage{graphicx}
\usepackage{natbib}
\usepackage{footnote}
\usepackage{multirow}
\usepackage{amsmath}
\usepackage{amssymb}
\usepackage{rotating}
\usepackage{hyperref}
\usepackage{xspace}
\usepackage{lineno}

\newcommand{\hii}{\ion{H}{2}\xspace}

\newcommand{\kms}{km s$^{-1}$\xspace}

\newcommand{\h}{$^{\mathrm{h}}$\xspace}
\newcommand{\m}{$^{\mathrm{m}}$\xspace}
\newcommand{\s}{$^{\mathrm{s}}$\xspace}
\defcitealias{Leroy18}{L18}

\shortauthors{Mills et al.}

\begin{document}

\title{Clustered Star Formation in the center of NGC 253 Contributes to Driving the Ionized Nuclear Wind}

\author{E. A. C. Mills}
\affiliation{Department of Physics and Astronomy, University of Kansas, 1251 Wescoe Hall Dr., Lawrence, KS 66045, USA}

\author{M. Gorski}
\affiliation{Department of Space, Earth and Environment, Astronomy and Plasma Physics, Chalmers University of Technology, SE-412 96, Gothenburg, Sweden}

\author{K.L. Emig}
\affiliation{Leiden Observatory, Leiden University, P.O.Box 9513, NL-2300 RA, Leiden, The Netherlands }
\affiliation{National Radio Astronomy Observatory,  520 Edgemont Rd, Charlottesville, VA 22903}
\altaffiliation{Jansky Fellow of the National Radio Astronomy Observatory}

\author{A. D. Bolatto}
\affiliation{Department of Astronomy and Joint Space-Science Institute, University of Maryland, College Park, MD 20742, USA}
\affiliation{Visiting Scholar, Flatiron Institute, Center for Computational Astrophysics, NY 10010, USA}

\author{R. C. Levy}
\affiliation{Department of Astronomy, University of Maryland, College Park, MD 20742, USA}

\author{A. K. Leroy}
\affiliation{Department of Astronomy, The Ohio State University, 140 West 18th Avenue, Columbus, OH 43210, USA}

\author{A. Ginsburg}
\affiliation{Department of Astronomy, University of Florida, 211 Bryant Space Sciences Center, Gainesville, 32611 FL, USA}

\author{J.D. Henshaw}
\affiliation{Max-Planck-Institut f\"{u}r Astronomie, K\"{o}nigstuhl 17, D-69117, Heidelberg, Germany}

\author{L. K. Zschaechner}
\affiliation{University of Helsinki, Physicum, Helsingin Yliopisto, Gustaf H\"{a}lstr\"{o}in katu 2, 00560 Helsinki, Finland}

\author{S. Veilleux}
\affiliation{Department of Astronomy and Joint Space-Science Institute, University of Maryland , College Park, MD 20742, USA}

\author{K. Tanaka}
\affiliation{Department of Physics, Faculty of Science and Technology, Keio University, 3-14-1 Hiyoshi, Yokohama, Kanagawa 223-8522 Japan}

\author{D. S. Meier}
\affiliation{New Mexico Institute of Mining \& Technology, 801 Leroy Place, Socorro, NM 87801, USA}
\affiliation{National Radio Astronomy Observatory, PO Box O, 1003 Lopezville Road, Socorro, New Mexico 87801, USA}

\author{F. Walter}
\affiliation{Max-Planck-Institut f\"{u}r Astronomie, K\"{o}nigstuhl 17, D-69117, Heidelberg, Germany}

\author{N. Krieger}
\affiliation{Max-Planck-Institut f\"{u}r Astronomie, K\"{o}nigstuhl 17, D-69117, Heidelberg, Germany}

\author{J. Ott}
\affiliation{National Radio Astronomy Observatory, PO Box O, 1003 Lopezville Road, Socorro, New Mexico 87801, USA}

\correspondingauthor{E.A.C Mills}
\email{eacmills@ku.edu}

\begin{abstract}
We present new 3 mm observations of the ionized gas toward the nuclear starburst in the nearby (D $\sim$3.5 Mpc) galaxy NGC 253. With ALMA, we detect emission from the H40$\alpha$ and He40$\alpha$ lines in the central 200 pc of this galaxy on spatial scales of $\sim$4 pc. The recombination line emission primarily originates from a population of approximately a dozen embedded super star clusters in the early stages of formation. We find that emission from these clusters is characterized by electron temperatures ranging from 7000-10000 K and measure an average singly-ionized helium abundance $\langle Y^+\rangle$= 0.25 $\pm$ 0.06, both of which are consistent with values measured for \hii regions in the center of the Milky Way. We also report the discovery of unusually broad-linewidth recombination line emission originating from seven of the embedded clusters. We suggest that these clusters contribute to the launching of the large-scale hot wind observed to emanate from the central starburst.  Finally, we use the measured recombination line fluxes to improve the characterization of overall embedded cluster properties, including the distribution of cluster masses and the fractional contribution of the clustered star formation to the total starburst, which we estimate is at least 50\%. 
\end{abstract}


\section{Introduction}

The barred spiral galaxy NGC 253 \citep[distance: 3.5 Mpc;][]{Rekola05} is one of the nearest examples of a galaxy undergoing a nuclear starburst. Apart from this key difference, the stellar component of the nucleus of NGC 253 is in many ways an analog to the Milky Way's center. Both are dominated by a bar potential and have a nearly identical stellar mass within the central 200 pc \citep[$\sim10^9$ M$_\odot$;][]{WynnWilliams79,Launhardt02,Sormani20b}. Additionally, while a supermassive black hole has not been definitively detected in NGC 253 \citep[and no AGN is present;][]{MullerSanchez10}, the virial mass in the central 20 pc is consistent with the mass of the Milky Way's central black hole \citep[$4-5\times10^6$ M$_\odot$;][]{RodriguezRico06,Ghez08}. However the star formation environments in the two nuclei are starkly different: within the central half kiloparsec of NGC 253, stars are forming at a rate of 2.8 M$_\odot$ yr$^{-1}$ \citep{Ott05,Bendo15}, fueled by a cold molecular gas reservoir with a mass of 2-4$\times10^8$ M$_\odot$\citep{Mauersberger96,Sakamoto11,Leroy15,Krieger19}. Compared to the center of the Milky Way, the center of NGC 253 has roughly an order of magnitude more molecular gas \citep{Dahmen98} and a 30-40 times higher star formation rate \citep{Longmore13,Barnes17}.

A significant fraction of the starburst activity in the central 200 pc of NGC 253 is concentrated in a population of embedded young massive clusters. \cite{Leroy18} (hereafter \citetalias{Leroy18}) detect 14 compact sources with continuum emission at both 33 GHz and 350 GHz, only one of which is visible at infrared wavelengths \citep{Watson96,Kornei09}. The large luminosities of these sources (from which stellar masses of $10^4-10^6$ M$_\odot$ are inferred), as well as the associated large gas masses of $\sim 5\times10^3-5\times10^5$ M$_\odot$, indicate that these sources are all in the process of forming super star clusters, which are typically defined as having initial stellar masses $> 10^5$ M$_\odot$ \citep{Turner00,Mengel02,Clark05}. \citetalias{Leroy18} further estimate that at least 20\% of the ionizing photons associated with the starburst come from the star formation in these clusters. They also suggest that star formation taking place in these and other undetected lower mass embedded clusters could account for the entire starburst.  Overall, the starburst in NGC 253  is relatively young.  Infrared spectra of the unobscured stellar population are consistent with ages of $<$ 8 Myr \citep{Kornei09,Davidge16}.  The deeply obscured cluster population is likely younger: ALMA observations by \citetalias{Leroy18} require they must be $\lesssim 5-10$ Myr.  They further argue that the theoretical time scale for the formation of clusters  still embedded in their natal gas suggests the embedded clusters should be as young as $\sim$1 Myr, and consistent with a zero-age main sequence (ZAMS) stellar population.

While the gas depletion time scale for the starburst nucleus of NGC 253 due to star formation alone is $\sim100$ Myr, the nucleus of NGC 253 also drives a massive molecular outflow detected with CO \citep[M$\sim10^7$ M$_\odot$;][]{Bolatto13,Zschaechner18,Krieger19} which is estimated to be additionally depleting the central reservoir at a rate of 10-40 M$_\odot$ yr$^{-1}$. Together with the star formation, this then suggests a gas depletion time scale of only $\sim$13-30 Myr. The outflow from NGC 253 is multiphase in nature \citep{Veilleux20} and is also detected in dense gas tracers like HCN $1-0$ \citep{Walter17} as well as H$_2$ $v=1-0$  S(1) \citep{Sugai03}, H$\alpha$ \citep{Sharp10,Westmoquette11}, and diffuse X-ray emission \citep{Pietsch00,Strickland02,Bauer07} with an extent of 100 pc to more than 1 kpc above the disk and deprojected speeds up to a few hundred \kms. However, the launching points of the outflow are not well localized beyond having an origin in the central 200 pc \citep{Bolatto13}, where it is believed to be driven purely by the starburst and not to be associated with AGN activity. 

As mentioned, although overall the nucleus of NGC 253 is gas rich, there is no clear candidate for an AGN. A bright nonthermal radio source \citep{TurnerHo85,Ulvestad97} has been suggested to be an embedded low-luminosity AGN, however it does not have the hard X-ray counterpart that would be expected in this case \citep{MullerSanchez10}, and so may instead be a supernova remnant. The location of a central supermassive black hole and true dynamical center of NGC 253 is thus uncertain; measurements of stellar kinematics suggest a location to within $\pm$15 pc, a region which includes the nonthermal radio source as well as an H$_2$O kilomaser \citep{Henkel04, Gorski19}, several hard x-ray sources \citep{MullerSanchez10}, and a source recently suggested to be the first extragalactic recombination line maser \citep{BaezRubio18}, while low-resolution H92$\alpha$ observations suggest an entirely different kinematic center \citep{AG96}. 

All together, the nucleus of NGC 253 is a complex environment made more difficult to interpret by its nearly edge-on orientation. Both embedded star formation and supernovae are seen in close (projected) proximity. Accurate determinations of the ionizing properties of the cluster stars from the continuum emission then requires disentangling the contributions from all of these potential sources. Despite these difficulties, the proximity of NGC 253 makes it the best laboratory for characterizing the dominant mode of star formation in nuclear starbursts. 

In this paper, we present new, high-resolution ($\sim 4$ pc) ALMA observations of the 3 mm continuum and the H40$\alpha$ line, both of which trace the ionized gas in the central few hundred parsecs of NGC 253. We use these data to reassess the cluster properties and their relation to the multiphase outflow. We also evaluate the recent claim that NGC 253 hosts the first detected extragalactic recombination line maser source \citep{BaezRubio18}.

\begin{figure*}
\includegraphics[scale=0.25]{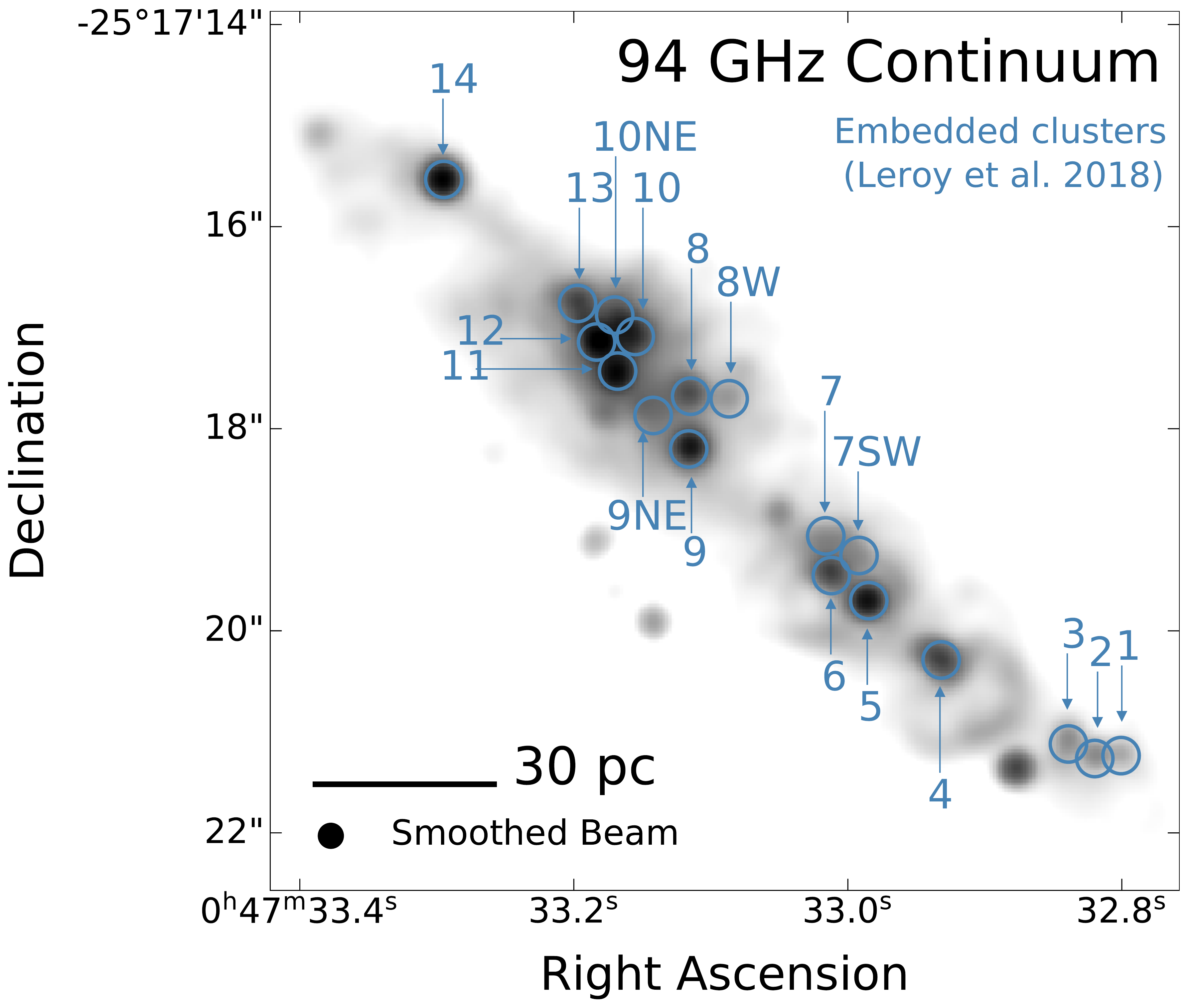}\includegraphics[scale=0.25]{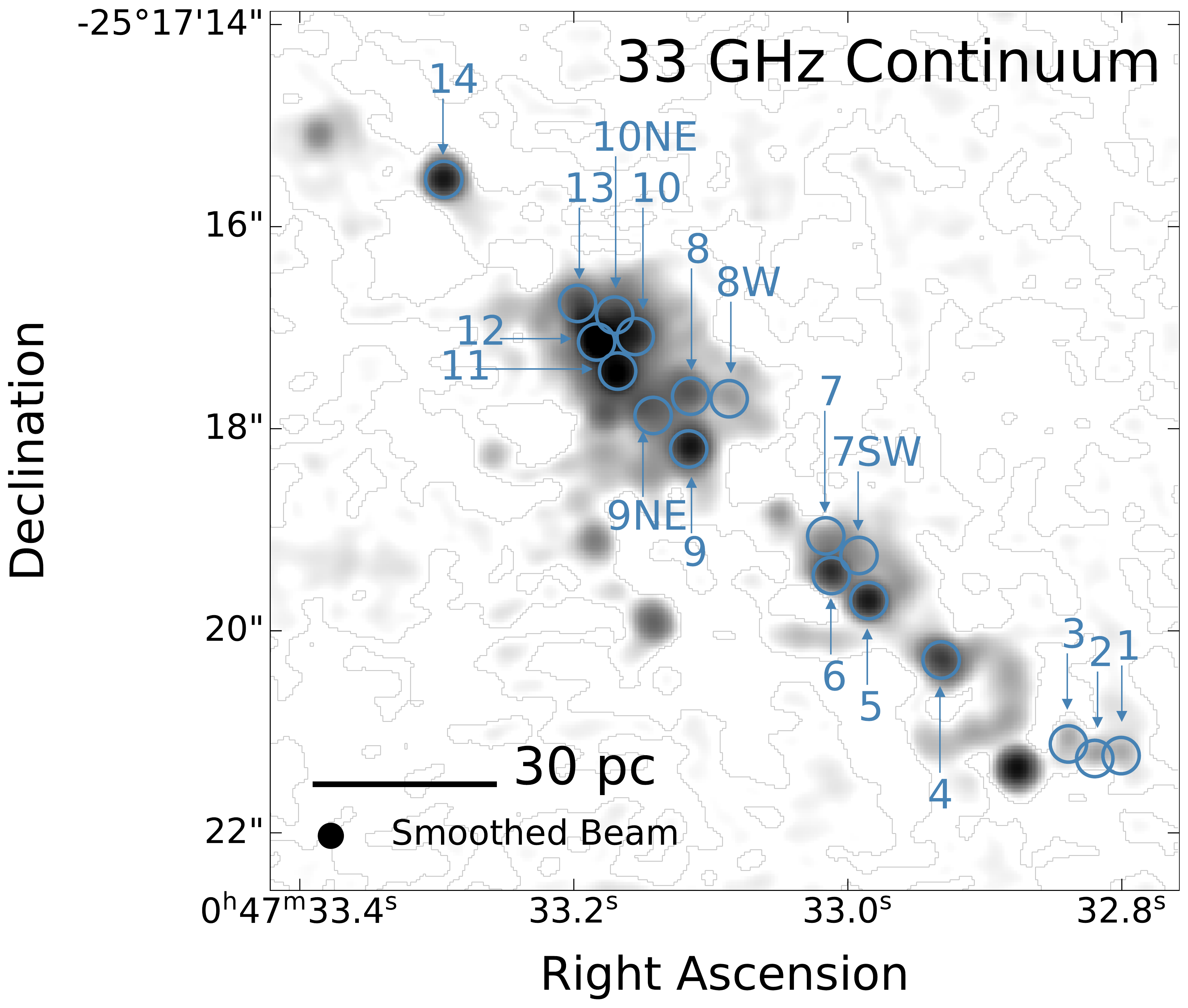} \\
\includegraphics[scale=0.25]{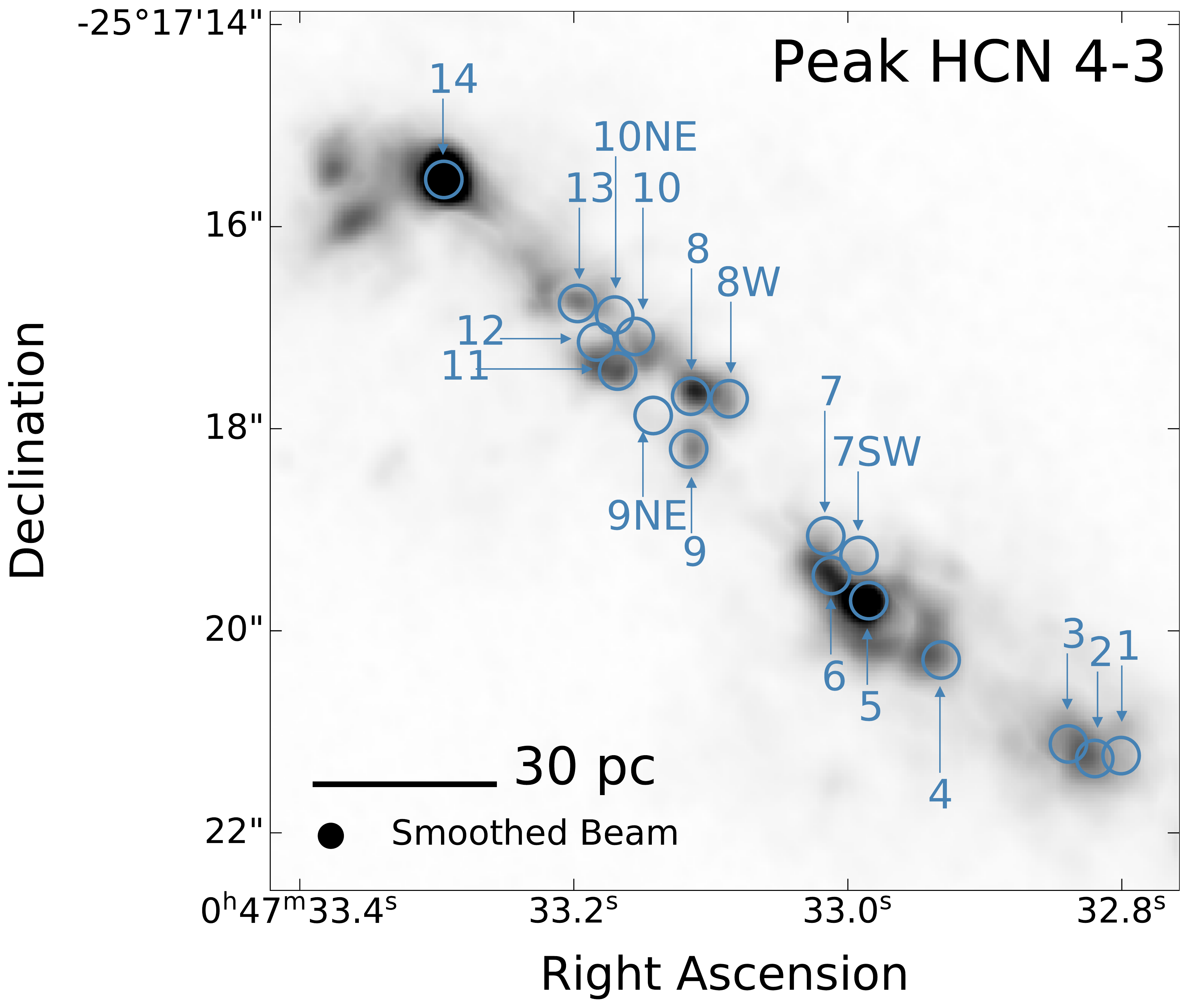}\includegraphics[scale=0.25]{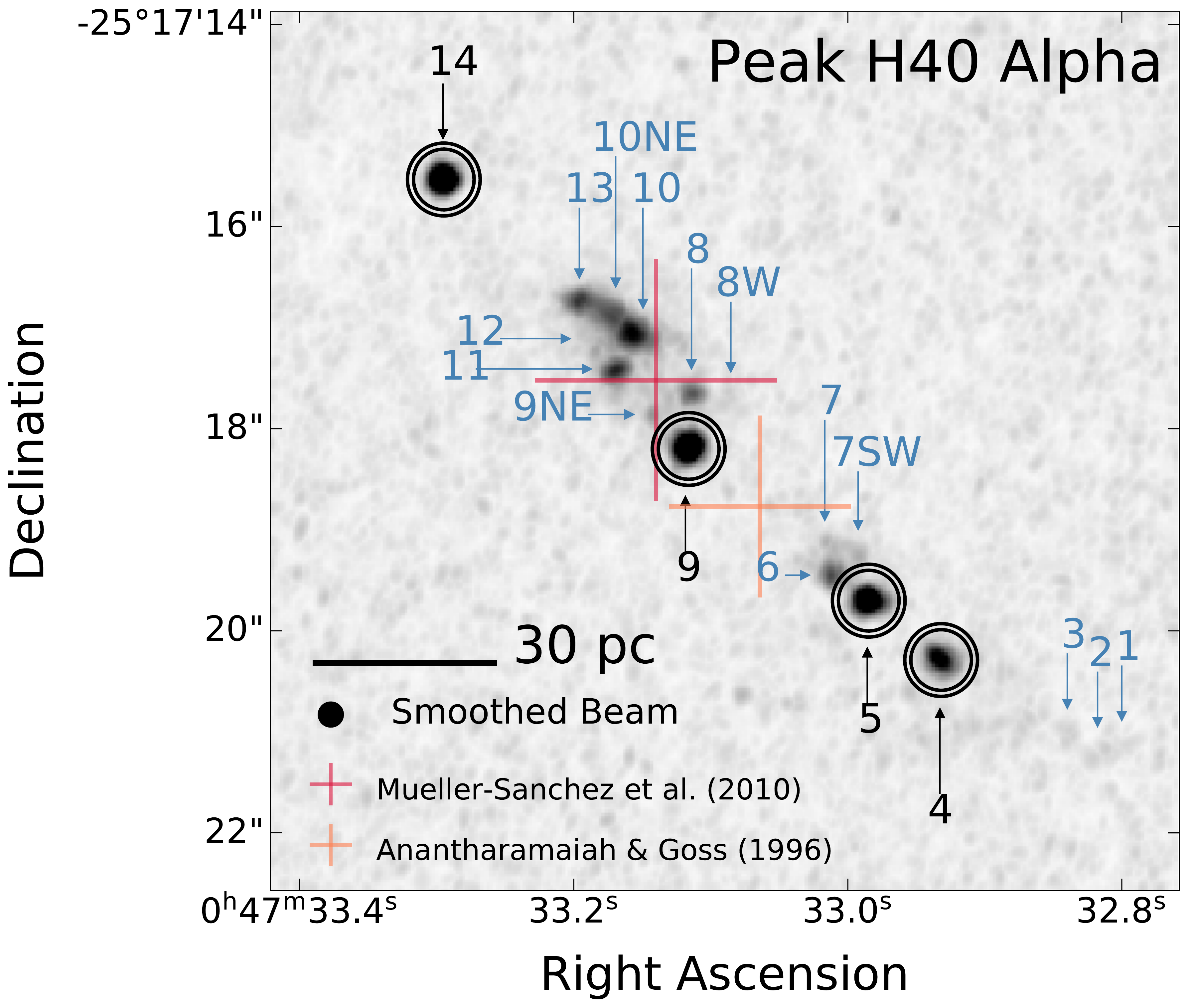} 
\caption{{\bf Top Left:} 94 GHz (3 mm) continuum emission toward the nucleus of NGC 253. Embedded cluster sources from \citetalias{Leroy18} are labeled with numbers 1-14; newly-identified sources have additional directional signifiers \citep[e.g., 10NE, as in ][]{BaezRubio18}. Circles indicate the $0\arcsec.18$ (3 pc) radius apertures used to extract recombination line properties. {\bf Bottom Left:} A map of the peak HCN 4$-$3 emission toward the nucleus of NGC 253 from \cite{Leroy18}. Labels are the same as for Top Left panel. {\bf Top Right:} 33 GHz continuum emission toward the nucleus of NGC 253 from \cite{Gorski17}. Labels are the same as for Top Left panel. {\bf Bottom Right:} A map of the peak H40$\alpha$ emission. Labels are the same as for Top Left panel. Circles indicate the larger apertures used to extract continuum properties for four isolated sources (4,5,9, and 14) in order to determine representative electron temperatures. The kinematic centers determined by \cite{MullerSanchez10} and \cite{AG96} and their 3$\sigma$ uncertainties are shown as crosses. } 
\label{Fig1}
\end{figure*}

\section{Observations}
\subsection{ALMA data}
The data used in this analysis were observed using the Atacama Large Millimeter and submillimeter Array (ALMA) in Cycle 5 (Project code 2017.1.00895.S, PI: E.A.C. Mills) in 15 sessions with an average of 45 antennas between November 25 and December 6, 2017. Observations were made in a single frequency setting at 3 mm (ALMA Band 3) toward a single pointing toward the nucleus of NGC 253 with a field of view (half power beam width) of $\sim1'$ ($\sim$1 kpc) in three extended configurations (C43-6, C43-7, and C43-8, with baselines ranging from 15-8500 m).

The frequency coverage of the data was chosen to match the first of two frequency settings observed at lower resolution ($\sim2"$, or $\sim$35 pc) in Cycle 0 with a 16 antenna array (Project code 2011.0.00172.S, PI: A. Bolatto) and subsequently published in \cite{Bolatto13}, \cite{Leroy15} and \cite{Meier15}. Four 1.875 GHz-wide subbands were centered on frequencies of 86.63 GHz, 88.48 GHz, 98.52 GHz and 100.38 GHz. The spectral resolution of these data is 0.977 MHz (2.8-3.4 km/s). This paper focuses on emission from jxust two lines, the H40$\alpha$ line at 99.022952 GHz and the He40$\alpha$ line at 99.063305 GHz. 

The calibration of the data was performed in the Common Astronomy Software Applications package (CASA\footnote{https://casa.nrao.edu/}) using the ALMA pipeline. Imaging of the continuum and the H40$\alpha$ and He40$\alpha$ lines for a single pointing was performed in CASA version 5.4.0 using the tclean task, with a robust weighting of 0.5 to optimize both resolution and sensitivity to the more extended line emission. The final images have beam sizes of $\sim0\arcsec.21\times0\arcsec.15$ ($3.6\times2.5$ pc) with $0\arcsec.03$ pixels and are sensitive to size scales up to $\sim$2$\arcsec$.6 ($\sim$44 pc) at 100 GHz, based on ALMA-provided statistics for the typical array configuration of these observations (C43-7). We smoothed both the line and continuum images to a circular beam with a common resolution of 0$\arcsec$.225 (3.8 pc). The RMS noise level for the continuum image is 7$\mu$Jy beam$^{-1}$, while the resulting RMS noise level in each $\sim$3\kms channel for the H40$\alpha$ data was $\sim$0.2 mJy beam$^{-1}$ (0.5 K). With almost 13 hours of on-source time on a single field, this is one of the deepest ALMA Band 3 images of any source to date. 

We also compare the Band 3 ALMA data to several other ALMA data sets covering the central region of NGC 253.  This includes Band 7 images of the HCN 4$-$3 line and the 345 GHz continuum  \citep[Project code 2015.1.00274.S; PI: A. Bolatto;][]{Leroy18}, which have a higher native resolution than the Band 3 data (0$\arcsec.$175 and 0$\arcsec.$11 respectively), and so are smoothed to the common resolution of 0$\arcsec$.225 (3.8 pc). We also include Band 6 continuum images at 240 GHz from the ALMA archive (Project code 2012.1.00789.S; PI: K. Nakanishi). These data have a lower angular resolution of $0\arcsec.5\times0\arcsec.4$, and so for analyses involving comparison with these data we smooth the Band 3 and Band 7 data to match this resolution. Both of these data sets are chosen as they trace the dense gas and dust surrounding the embedded clusters. The final data set we compare to is archival H26$\alpha$ line data \citep[Project code 2013.1.00735.S; PI: K. Nakanishi;][]{BaezRubio18}. These data also have a somewhat lower angular resolution of $\sim0\arcsec.3$, and so for analyses including these data we smooth our H40$\alpha$ data to a matching resolution.  

\subsection{VLA data}
The ALMA data are additionally compared to data from the Karl G. Jansky Very Large Array (VLA)\footnote{The National Radio Astronomy Observatory is a facility of the National Science Foundation operated under cooperative agreement by Associated Universities, Inc.}.
The VLA data consist of 22 and 33 GHz continuum emission maps toward the center of NGC 253 \citep{Gorski17,Gorski19}. These data have native resolutions that are higher than the ALMA images, with a FWHM beam at 22 (33) GHz of 0$\arcsec$.204$\times$0.$\arcsec$121 (0$\arcsec$.096$\times$0.$\arcsec$045), and are sensitive to emission on spatial scales less than 2.$\arcsec$4 (1.$\arcsec$6). We convolve them to a 0.$\arcsec$225 beam to match the resolution of the ALMA data for this analysis, and regrid them to an identical pixelization. After this convolution the 22 (33) GHz VLA data have a continuum RMS noise of 90 (40) $\mu$Jy beam$^{-1}$.

We find that the VLA and ALMA reference frames are not exactly aligned. While there is no strong reason to prefer one reference frame to the other, we note that positional errors have been seen in other high-frequency observations made at low elevation with the VLA \citep{Mills18c}, and that these VLA data in particular were taken at a time when the proper refraction calibration was not being applied\footnote{https://science.nrao.edu/facilities/vla/data-processing/vla-atmospheric-delay-problem}. We thus apply a slight shift to the VLA data to align it with the positional reference frame of the ALMA images. We perform a cross correlation between the ALMA continuum image and the VLA images using the task `register\_translation' from the python package scikit-image \citep{scikit14}, and shift the VLA images by $0\arcsec.035$ in Right Ascension and $0\arcsec$ in Declination. The uncertainty in the alignment from the cross correlation is $\pm0\arcsec.012$, an order of magnitude smaller than the size scales of the sources studied in this work. 

\section{Results}

The 94 GHz (Band 3) continuum map is shown in the top left panel of Figure \ref{Fig1}. This image reveals more than two dozen sources of compact emission as well as significant extended emission structures, including ridges, diffuse emission, and shells. For comparison, we also show in Figure \ref{Fig1} the map of peak H40$\alpha$ recombination line emission (bottom right) and previously-published maps of HCN 4-3 emission \citepalias[bottom left;][]{Leroy18} and 33 GHz continuum emission \citep[top right;][]{Gorski19}

\begin{figure}
\includegraphics[scale=0.21]{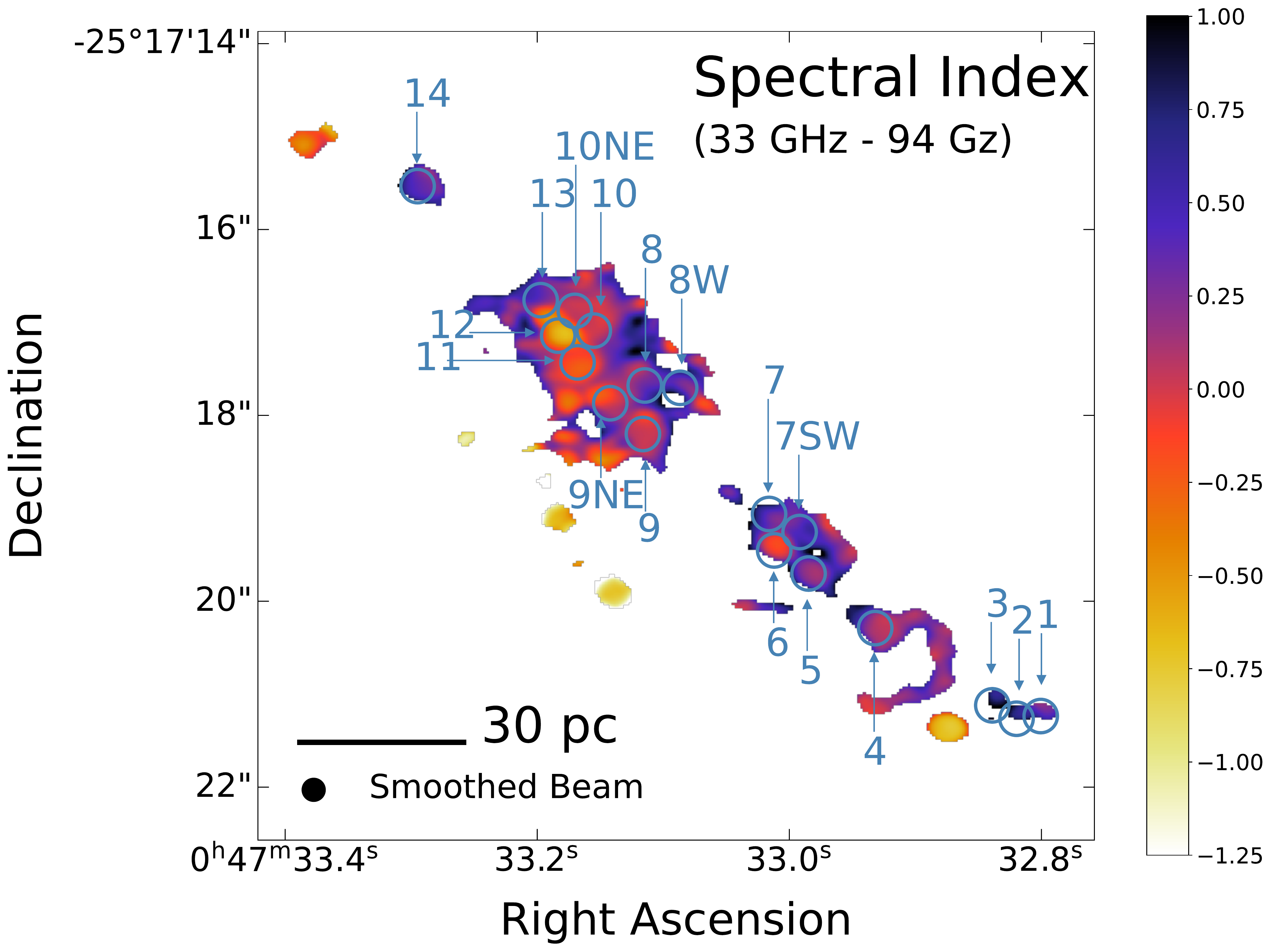}
\caption{Spectral index map of emission between 33 GHz and 94 GHz.  Sources are labeled as in Figure \ref{Fig1}.  While many sources have a superposition of emission from different mechanisms, sources that are most likely dominated by nonthermal emission have negative spectral indices $<-0.5$. Sources that are predominately thermal free-free emission are expected to have spectral indices near 0, while sources with spectral indices $>0.5$ likely have some contribution from dust emission.}
\label{Fig2}
\end{figure}

\subsection{Identification of Free-Free Emission Sources}
We expect that for many locations in the actively star-forming nucleus of NGC 253, there may be a superposition of different types of emission mechanisms (e.g., thermal free-free, nonthermal synchrotron, and thermal dust) associated with different stages of massive star evolution (e.g., supernovae, embedded clusters, and molecular clouds). We use the H40$\alpha$ emission to identify  sources of free-free emission corresponding to embedded young clusters, as this line emission originates solely in the ionized gas surrounding these clusters. We note that the presence of H40$\alpha$ does not indicate an absence of nonthermal or dust emission, and the identified clusters may still have significant nonthermal or dust contributions to their 94 GHz continuum fluxes. We indicate in Figure \ref{Fig1} the location of all of the embedded young clusters identified by \citetalias{Leroy18} (labeled with numbers 1-14), however we update the position for several of these sources (4, 5, 6, 7, 10, 11, 12, and 14) as they were not well-centered on the free-free emission, likely due to physical offsets between the dust continuum and the radio continuum. We take the location of the free-free emission to be more representative of the embedded cluster location. We further identify several additional sources of compact thermal emission that were not identified by \citetalias{Leroy18}, which we designate as 7SW, 9NE, 8W and `10NE' (the latter of which was previously identified by \citealt{BaezRubio18}). While the H40$\alpha$ line emission is strongest from sources 4, 5, 6, 8, 9, 10, 10NE, 11, 12, 13 and 14, recombination line emission is detected toward all of the identified sources, as we discuss in the following sections.   

\subsubsection{Nonthermal Sources}
In addition to the sources labelled in Figure \ref{Fig1}, there are a number of strong sources of compact emission visible in the 33 GHz and 94 GHz continuum images, primarily to the south of the main band of identified sources.  These sources are most likely dominated by synchrotron emission, as can be seen in a spectral index map ($S_\nu \propto \nu^{\alpha}$) calculated between the 33 and 94 GHz continuum images (Figure \ref{Fig2}). The likely synchrotron-dominated sources exhibit spectral-indices $\alpha < -0.5$. Their dominant nonthermal nature is supported by the lack of detectable emission in either HCN 4$-$3 or H40$\alpha$.  All of these sources are also prominent in 8 and 15 GHz continuum images from \cite{Ulvestad97}, supporting the interpretation that these are nonthermal sources, likely supernova remnants.

\subsection{H40\texorpdfstring{$\alpha$}{-alpha} Recombination Line Emission}
\label{recomb}

\begin{figure}
\includegraphics[scale=0.37]{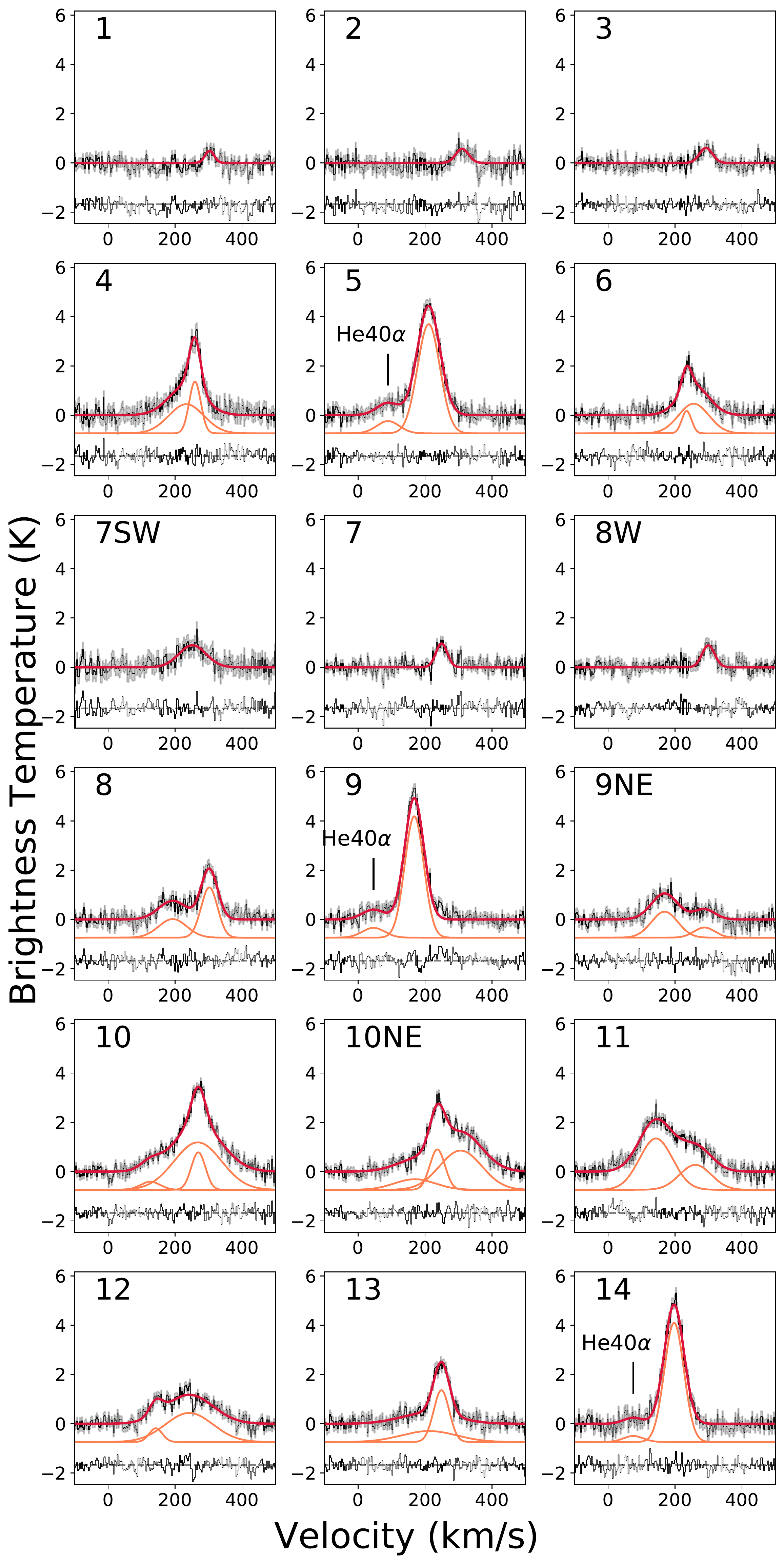}
\caption{Spectra of H40$\alpha$ toward all sources labeled in Figure \ref{Fig1}. One, two, and three-component Gaussian fits are conducted to account for gas at different line-of-sight velocities, and the resulting fit for each spectrum is plotted in red. For sources where there are multiple components fit, the individual Gaussian components are also plotted in orange. Residuals from the fitting are plotted below the spectra. The results of these fits are given in Table \ref{params}. The location of the He40$\alpha$ line is indicated in the three sources (5, 9 and 14) where it is clearly detected.}
\label{Fig3}
\end{figure}

\begin{figure}
\includegraphics[scale=0.37]{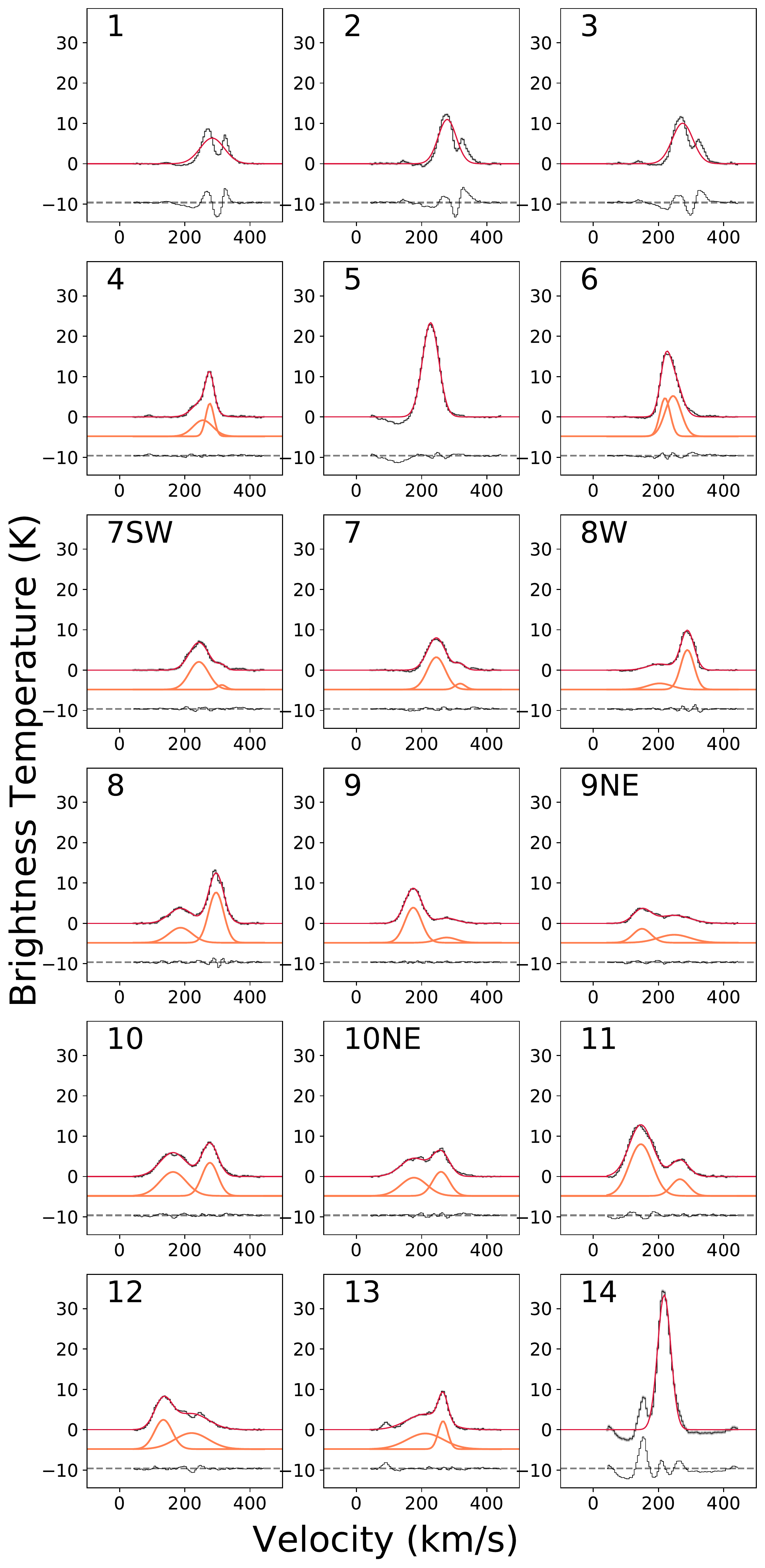}
\caption{Spectra of HCN 4$-$3 toward all sources labeled in Figure \ref{Fig1}. The velocity axis is chosen to cover the same range as the H40$\alpha$ spectra. One and two-component Gaussian fits are conducted, and the best-fit model is overplotted in red. Where two Gaussian components are fit, we show the individual components in orange. The residuals of these fits are plotted below the spectra. The results of these fits are given in Table \ref{kinematics}.}
\label{Fig4}
\end{figure}

\begin{figure}
\includegraphics[scale=0.26]{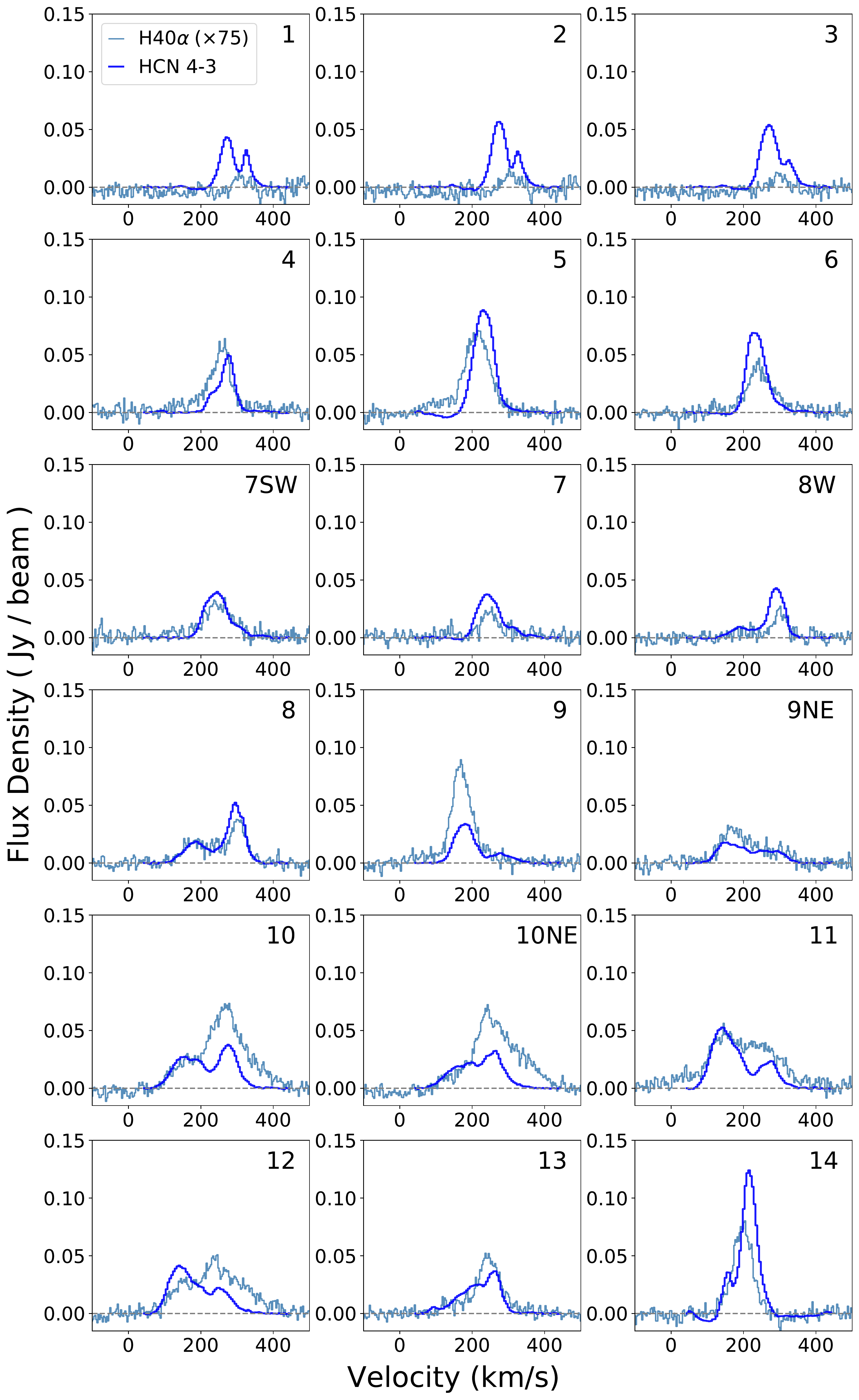}
\caption{Spectra of H40$\alpha$ (light blue) and HCN 4$-$3 (dark blue) toward all sources labeled in Figure \ref{Fig1}. The H40$\alpha$ spectra are scaled by a factor of 75 in order to facilitate comparison.}
\label{Fig5}
\end{figure}

Spectra of the H40$\alpha$ line toward all sources are extracted from a r=$0\arcsec.18$ (3 pc) aperture and shown in Figure \ref{Fig3}. We note that there is a slight overlap in the apertures centered on sources 1, 2, and 3. We attempt to quantify the effects of this overlap by measuring the amount of 3 mm continuum emission present in the overlap region. Sources 1 and 2 have 20-25\% of their total flux in the overlap area, implying that $\sim$10\% of the emission in the aperture of each source may be due to contamination from the other source. In contrast, sources 2 and 3 have only 7\% of their total flux in the overlap area, implying $<$5\% contamination. We thus do not expect these aperture overlaps to affect our results. We conduct Gaussian line fitting for all sources using PySpecKit \citep{Ginsburg11}. For sources 4, 6, 8, 9NE, 10, 10NE, 11, 12, and 13 we fit 2-3 separate Gaussian profiles to the H40$\alpha$ line emission to account for the presence of multiple velocity components along the line of sight. The results of the H40$\alpha$ line fits are given in Table \ref{params}.

We also conduct Gaussian fits to the HCN 4$-$3 line profiles extracted from the same aperture, shown in Figure \ref{Fig4}. This line traces dense gas associated with the embedded clusters, and exhibits many kinematic similarities with the ionized gas traced by H40$\alpha$. Similar to the H40$\alpha$ fitting, we fit for multiple Gaussian components in sources 4, 6, 7, 7SW, 8, 8W, 9, 9NE, 10, 10NE, 11, 12 and 13. We do not fit for multiple components in sources 1, 2, 3 and 14, as these profiles all exhibit a central dip at the same velocity as the H40$\alpha$ emission. This dip is likely not due to the presence of multiple velocity components, but instead due to HCN self-absorption, which is also seen on comparable spatial scales in lower-excitation lines of multiple species in the embedded cluster Sgr B2\footnote{Sgr B2 is likely the best local analog for the embedded cluster sources we observe in NGC 253. The estimated zero-age main sequence stellar mass for Sgr B2 \citep[20,000-45,000 M$_\odot$;][]{Schmiedeke16,Ginsburg18a} is comparable to the stellar masses inferred for some of the lowest-mass sources seen here, like the heavily embedded sources 1, 2, and 3, and its line widths are similar \citep{Leroy18}.} in the center of the Milky Way \citep{Mills17a}. As a result, these fits do not yield a good estimate of the line widths, and we see large residuals left after fitting. Additionally, we also see some residual absorption at low velocities in sources 5 and 14, which could be a P-cygni profile, indicative of outflowing gas \citep{Levy21}. Profiles of the HCN 4$-$3 line overlaid on the H40$\alpha$ line are shown in Figure \ref{Fig5}. A comparison of the kinematics of the H40$\alpha$ and HCN 4$-$3 lines for each source is given in Table \ref{kinematics}. 

Measured H40$\alpha$ line widths ($\sigma_{FWHM}$) range from 40-100 \kms for most sources. However toward sources 4, 6, 10, 10NE, 11, 12, and 13 we measure a much broader linewidth component ($\Delta v_{FWHM}$ = 100-200 \kms). This component is $\sim$ 50 \kms\, broader than the HCN 4$-$3 emission located at a comparable velocity centroid. In addition, we detect a broad-linewidth component in sources 10 and 10 NE at large positive velocities (270-300 \kms) that has no HCN counterpart. For both of these components we investigate the possibility of line contamination, but find no plausible identification for this feature apart from H40$\alpha$.  Several other sources (5, 7SW, 8, 9NE, and 14) also have H40$\alpha$ emission that is 10-30 \kms broader than the HCN 4$-3$ emission. In contrast, sources 1, 2, 3, and 7 exhibit narrower H40$\alpha$ emission than HCN 4$-$3 emission. While the HCN linewidth in sources 1, 2, and 3 is likely underestimated due to self-absorption, a narrower H40$\alpha$ line profile in source 7 could occur if this is a less evolved source, in which there is a lower stellar mass in the inner region of the central forming cluster compared to the surrounding molecular gas reservoir at larger radii traced by HCN \citep[as discussed in Section 4.3 of ][]{Leroy18}. We discuss the gas kinematics further, including possible origins for the broad-linewidth ionized gas component, in Section \ref{dis}. 

\subsubsection{He40\texorpdfstring{$\alpha$}{-alpha} Emission and Ionized Helium Fraction}

For several sources we also detect a weaker emission feature corresponding to the He40$\alpha$ line, which is offset by $-122.2$ \kms\ from the H40$\alpha$ line. We do not, however, detect any carbon recombination line emission (the C40$\alpha$ line is offset by -150 \kms\, from H40$\alpha$) toward any source. For sources 5, 9, and 14, we have a clear detection of the He40$\alpha$ line, which can be seen in Figure \ref{Fig3}. These three sources are the strongest H40$\alpha$ emitters, and an unambiguous identification of He40$\alpha$ is possible due to the simple (single-component) structure of their velocity profiles. There are also several sources that have an excess of low-velocity emission that could be due to He40$\alpha$, but a clear identification is not possible. In sources 4, 8 and 13, a possible He40$\alpha$ feature overlaps with a lower-velocity gas component seen in HCN 4$-$3, and in source 11 the broad, multi-component nature of the velocity profile together with a lack of HCN 4-3 data for this velocity range make the identification of He40$\alpha$ emission ambiguous. For the three sources with clear He40$\alpha$ emission, we simultaneously fit two independent Gaussian profiles to both recombination lines, and report the results of these fits in Table \ref{params}.  

The simultaneous detection of H40$\alpha$ and He40$\alpha$ emission can be used to determine the helium abundance in these three sources, an indication of enrichment from prior generations of star formation. For the relatively high $n$ states of hydrogen and helium probed by radio/millimeter recombination lines, the level populations should be affected by radiative and collisional effects in the same way. As a result, the ratio of the line areas will then be equal to the ionic abundance ratio \citep{Churchwell74,Mezger76}. The mass-weighted fraction of singly-ionized $^4$He ($Y^+$) is then determined as 

\begin{equation}
Y^+ =   \frac{\int T_{B}[\mathrm{He}]\mathrm{d}v\ M_{\mathrm{He}}}{\int T_{B}[\mathrm{He}]\mathrm{d}v\ M_{He} + \int T_{B}[H]\mathrm{d}v\ M_{H}}.
\end{equation}

Here, $\int T_{B}[\mathrm{He}]$d$v$ is the integrated line intensity of the He40$\alpha$ line, $M_{He}$ is the mass of $^4$He, $\int T_{B}[\mathrm{H}]$d$v$ is the integrated line intensity of the H40$\alpha$ line (all integrations are done in the velocity domain), and $M_{H}$ is the mass of H. The measured $Y^{+}$ approaches the mass-weighted abundance of singly-ionized helium ($Y$) if the fraction of doubly-ionized helium is neglible \citep[this is generally true for Galactic HII regions;][]{Churchwell74} and if the radiation field in these sources is strong enough to fully singly-ionize the helium, making the volume-emitting regions of He$^+$ and H$^+$ the same \citep[this should be the case for stars with spectral type O9 and earlier, and is assumed to be a valid assumption for the clusters here;][]{Mezger74,Churchwell74}. We also look for emission from the HeII (64) $\alpha$ line at 98.0795642 GHz, but do not detect any.  Models predict that in the presence of an AGN, recombination lines from He$^{+}$ would be stronger than the hydrogen recombination lines \citep{Scoville13}, thus our nondetection is consistent with other evidence that NGC 253 does not host a central AGN. 

We report the derived $Y^+$ values in Table \ref{params}. For sources 5 and 9 we measure $Y^+$ values $\sim0.3$, while for source 14 we measure $Y^+ \sim0.17$. An abundance $ Y^+\sim0.3$ is consistent with He$^+$/H$^+$ values measured in the same way for \hii regions in the Milky Way's center, using lower-frequency (3.6 cm - 7 mm) recombination lines. Values in the Milky Way center range from 0.1-0.3 in the Arched Filaments and Sgr A West \citep[][though these early measurements may have been systematically biased to low values]{Mezger76} and up to 0.2-0.45 in the analogous massive protocluster Sgr B2 \citep{dePree96} and 0.39 in the nebula surrounding the luminous blue variable Pistol Star \citep{Lang97}. As in the Galactic center, the relatively high ratios of He$^+$ to H$^+$ observed in the NGC 253 sources indicate the presence of early O-stars and/or Wolf Rayet stars. Values toward the Milky Way center are also consistent with though somewhat higher than average values in the Milky Way disk \citep[$\langle Y^+_{disk}\rangle = 0.21\pm 0.08$;][]{Wenger13}, which would be expected based on measurements of an increasing Helium abundance with decreasing galactocentric radius \citep{MendezDelgado20}. While the lower value seen in source 14 could be an indication of a lower helium abundance in this source, we view it as more likely that this is an underestimate. If the He Str\"{o}mgren sphere is both smaller than the H Str\"{o}mgren sphere and the resolution of the observations (as might be expected if the radiation field is not as hard as in other sources), the He40$\alpha$ line will have a larger beam dilution factor, leading to an underestimate of the true helium abundance. 

\setlength{\tabcolsep}{4pt}

\begin{table*}[t]
\caption{Recombination Line Parameters}
\centering
\begin{tabular}{llllllllll}
\hline\hline
       &  RA       & Dec       & \multicolumn{3}{c}{----------- H40 $\alpha $ -----------}                 & \multicolumn{3}{c}{----------- He40 $\alpha $ -----------}                   & Y$^+$ \\
ID & ($\degr$) & ($\degr$) & v$_{cen}$    &  $\sigma_{\mathrm{FWHM}}$ & Peak T$_B$ &  v$_{cen}$   & $\sigma_{\mathrm{FWHM}}$ & Peak T$_B$ &        \\
       &           &           &(km s$^{-1}$) & (km s$^{-1}$)          & (K)        &(km s$^{-1}$) & (km s$^{-1}$)             & (K)        &        \\
\hline
1& 11.88667 & -25.28923 & 303.1 $\pm$ 2.5 & 32 $\pm$ 5 & 0.5 $\pm$ 0.1 & & & $<$0.18 & $<$0.60  \\
2& 11.88675 & -25.28924 & 310.7 $\pm$ 3.3 & 45 $\pm$ 7 & 0.6 $\pm$ 0.1 & & & $<$0.24 & $<$0.62  \\
3& 11.88683 & -25.28920 & 292.3 $\pm$ 2.2 & 45 $\pm$ 5 & 0.6 $\pm$ 0.1 & & & $<$0.17 & $<$0.52  \\
4& 11.88722 & -25.28897 & 233.5 $\pm$ 4.8 & 123 $\pm$ 8 & 1.2 $\pm$ 0.1 & & & $<$0.16 & $<$0.42  \\
& & & 259.3 $\pm$ 1.1 & 40 $\pm$ 3 & 2.1 $\pm$ 0.2 & &  \\
5&  11.88744 & -25.28881 & 210.9 $\pm$ 0.5 & 78 $\pm$ 1 & 4.4 $\pm$ 0.1 & 211.5 $\pm$ 0.5 & 79 $\pm$ 12 & 0.50 $\pm$ 0.06 & 0.31 $\pm$ 0.06 \\
6& 11.88755 & -25.28874 & 234.7 $\pm$ 1.6 & 34 $\pm$ 4 & 0.9 $\pm$ 0.1 & & & $<$0.18 & $<$0.79  \\
& & & 255.7 $\pm$ 2.8 & 106 $\pm$ 5 & 1.2 $\pm$ 0.1 & &  \\
7SW& 11.88747 & -25.28868 & 251.8 $\pm$ 3.9 & 95 $\pm$ 9 & 0.9 $\pm$ 0.1 & & & $<$0.24 & $<$0.52  \\
7& 11.88757 & -25.28863 & 249.3 $\pm$ 1.2 & 40 $\pm$ 2 & 1.0 $\pm$ 0.1 & & & $<$0.18 & $<$0.63  \\
8W& 11.88786 & -25.28825 & 299.5 $\pm$ 1.7 & 44 $\pm$ 3 & 0.9 $\pm$ 0.1 & & & $<$0.19 & $<$0.45  \\
8& 11.88798 & -25.28824 & 193.0 $\pm$ 3.6 & 98 $\pm$ 9 & 0.8 $\pm$ 0.0 & & & $<$0.19 & $<$0.50  \\
& & & 302.3 $\pm$ 1.0 & 57 $\pm$ 2 & 2.0 $\pm$ 0.1 & &  \\
9&  11.88798 & -25.28839 & 167.8 $\pm$ 0.4 & 66 $\pm$ 0 & 4.9 $\pm$ 0.1 & 168.3 $\pm$ 0.5 & 79 $\pm$ 12 & 0.40 $\pm$ 0.05 & 0.28 $\pm$ 0.05 \\
9NE& 11.88809 & -25.28830 & 168.8 $\pm$ 2.4 & 92 $\pm$ 6 & 1.1 $\pm$ 0.0 & & & $<$0.18 & $<$0.41  \\
& & & 287.9 $\pm$ 5.6 & 73 $\pm$ 13 & 0.4 $\pm$ 0.1 & &  \\
10& 11.88815 & -25.28808 & 126.1 $\pm$ 5.4 & 68 $\pm$ 13 & 0.3 $\pm$ 0.1 & & & $<$0.14 & $<$0.63  \\
& & & 269.5 $\pm$ 1.0 & 49 $\pm$ 3 & 1.5 $\pm$ 0.1 & &  \\
& & & 267.9 $\pm$ 2.0 & 166 $\pm$ 7 & 1.9 $\pm$ 0.1 & &  \\
10NE& 11.88821 & -25.28802 & 169.3 $\pm$ 33.6 & 142 $\pm$ 44 & 0.4 $\pm$ 0.1 & & & $<$0.15 & $<$0.57  \\
& & & 236.8 $\pm$ 1.2 & 53 $\pm$ 4 & 1.6 $\pm$ 0.2 & &  \\
& & & 306.3 $\pm$ 9.0 & 147 $\pm$ 11 & 1.6 $\pm$ 0.1 & &  \\
11& 11.88820 & -25.28817 & 143.1 $\pm$ 3.6 & 116 $\pm$ 5 & 2.1 $\pm$ 0.1 & & & $<$0.17 & $<$0.24  \\
& & & 260.8 $\pm$ 7.0 & 110 $\pm$ 11 & 1.0 $\pm$ 0.1 & &  \\
12& 11.88826 & -25.28809 & 142.4 $\pm$ 2.4 & 45 $\pm$ 7 & 0.6 $\pm$ 0.1 & & & $<$0.16 & $<$0.53  \\
& & & 242.1 $\pm$ 3.6 & 166 $\pm$ 7 & 1.2 $\pm$ 0.0 & &  \\
13& 11.88832 & -25.28799 & 212.1 $\pm$ 9.3 & 193 $\pm$ 18 & 0.4 $\pm$ 0.1 & & & $<$0.18 & $<$0.61  \\
& & & 248.9 $\pm$ 0.8 & 54 $\pm$ 2 & 2.1 $\pm$ 0.1 & &  \\
14&  11.88873 & -25.28765 & 197.2 $\pm$ 0.5 & 68 $\pm$ 1 & 4.8 $\pm$ 0.1 & 197.8 $\pm$ 0.5 & 69 $\pm$ 22 & 0.25 $\pm$ 0.07 & 0.17 $\pm$ 0.07 \\
\hline\hline
\end{tabular}
\label{params}
\end{table*}

\subsubsection{Investigating a Proposed Recombination Line Maser in Source 10NE}

\begin{figure}
\includegraphics[scale=0.5]{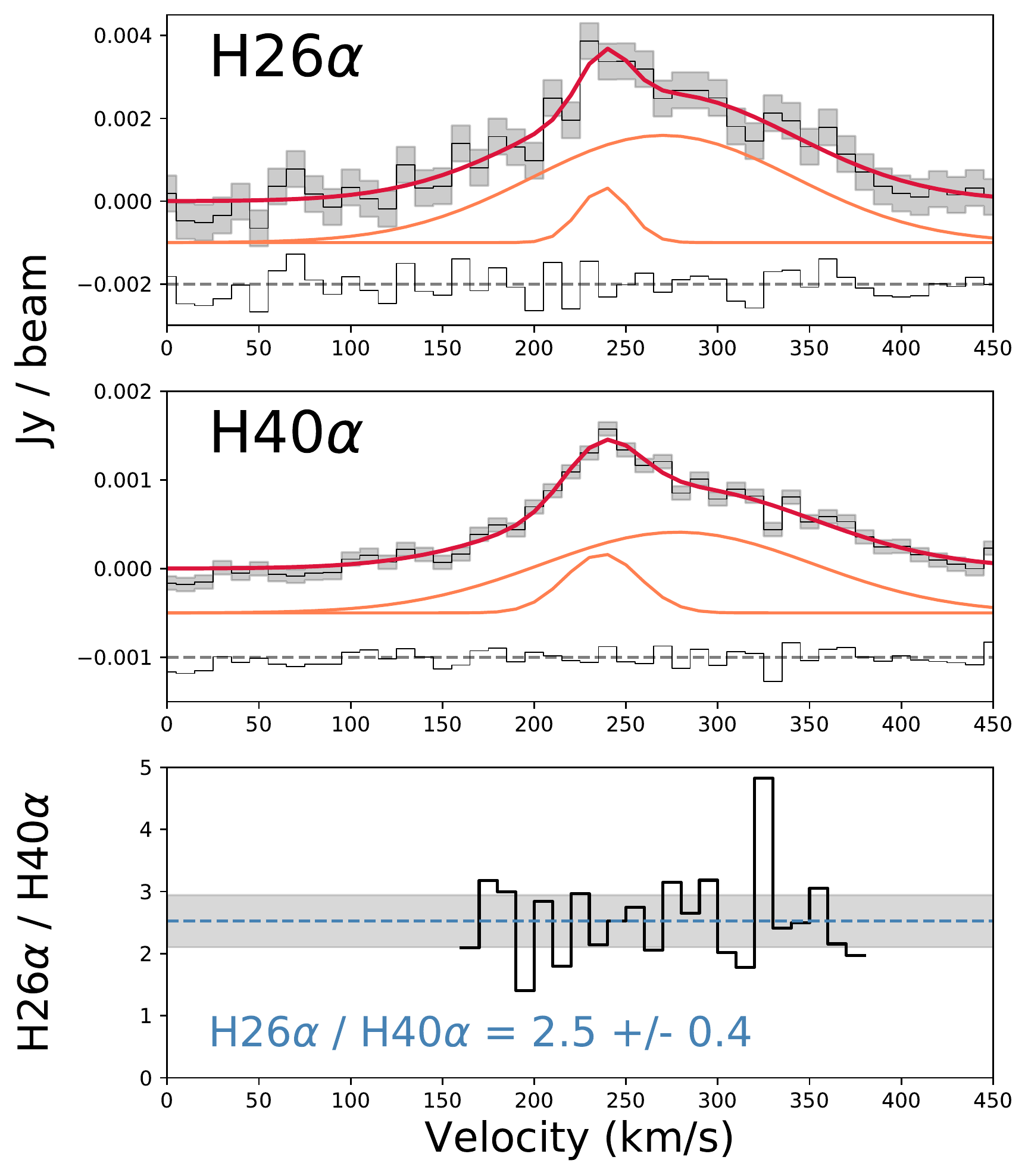}
\caption{{\bf Top and Middle:} Spectra of H26$\alpha$ and H40$\alpha$ in a r=$0''.18$ aperture toward source `10NE'. Two-component Gaussian fits are conducted for both lines and the best-fit model is overplotted in red. Individual components of these fits are shown below the spectra in orange. The residuals of these fits are also plotted below the spectra. {\bf Bottom:} The ratio of the two lines as a function of velocity. The mean value of this ratio is plotted as a light blue dotted line. We do not see the excess of emission in the H26$\alpha$ line at a velocity around 260 km/s that was claimed by
\cite{BaezRubio18}.}
\label{Fig6}
\end{figure} 

For source 10NE, we compare the H40$\alpha$ line in our data with archival images of the H26$\alpha$ line at 353.622747 GHz ($0\arcsec.3$ resolution, from project 2013.1.00735.S, PI: K. Nakanishi). Smoothing the H40$\alpha$ line to a matching $0\arcsec.3$ resolution and extracting spectra from a r=$0\arcsec.18$ aperture centered on 10NE, we find a consistent flux ratios between the H26$\alpha$ and H40$\alpha$ line of 2.5$\pm$0.5, as shown in Figure \ref{Fig6}. This corresponds to a recombination line spectral index $\alpha_L=0.73\pm0.13$. This is slightly smaller than the expectation for simple, optically-thin emission ($\alpha_L=1$), and is consistent with expectations for thermal emission \citep{Prozesky20}. It is not consistent with the $\alpha_L>2.1$ measured between the H26$\alpha$ and H30$\alpha$ line, which was argued to be evidence for stimulated emission in this line \citep{BaezRubio18}. We therefore find no evidence for recombination line maser emission toward this source. 

\begin{table*}[t]
\caption{Kinematic Comparison of Ionized and  Molecular Gas}
\centering
\begin{tabular}{lllllll}
\hline\hline
       & \multicolumn{2}{c}{H40$\alpha$ } &\multicolumn{2}{c}{HCN $(4-3)$} & $\Delta$ v$_{cen}$  & $\Delta$ v$_{FWHM}$ \\
Source &    v$_{cen}$    &  v$_{FWHM}$   & v$_{cen}$    &  v$_{FWHM}$                      & (H40$\alpha$ - HCN)  & (H40$\alpha$ - HCN)\\
       &   (km s$^{-1}$) & (km s$^{-1}$) &(km s$^{-1}$) & (km s$^{-1}$)                    & (km s$^{-1}$)       & (km s$^{-1}$) \\
\hline
1& 303.1 $\pm$ 2.5 & 32.3 $\pm$ 5.9 & 283.9 $\pm$ 0.2 & 87.3 $\pm$ 0.5 & 19.2 & -55.1  \\
2& 310.7 $\pm$ 3.3 & 45.6 $\pm$ 7.8 & 278.5 $\pm$ 0.2 & 65.1 $\pm$ 0.5 & 32.2 & -19.5  \\
3& 292.3 $\pm$ 2.2 & 45.8 $\pm$ 5.2 & 274.1 $\pm$ 0.2 & 74.5 $\pm$ 0.5 & 18.2 & -28.7  \\
4& 233.5 $\pm$ 4.8 & 123.3 $\pm$ 8.6 & 255.5 $\pm$ 1.2 & 70.9 $\pm$ 1.4 & -22.0 & 52.4  \\
&  259.3 $\pm$ 1.1 & 40.5 $\pm$ 3.5 & 276.6 $\pm$ 0.2 & 29.7 $\pm$ 0.7 & -17.3 & 10.8  \\
5& 210.9 $\pm$ 0.5 & 78.7 $\pm$ 1.3 & 227.5 $\pm$ 0.0 & 59.0 $\pm$ 0.1 & -16.6 & 19.7  \\
6& 234.7 $\pm$ 1.6 & 34.1 $\pm$ 4.8 & 220.2 $\pm$ 0.3 & 37.2 $\pm$ 1.1 & 14.5 & -3.1  \\
&  255.7 $\pm$ 2.8 & 106.6 $\pm$ 5.2 & 244.9 $\pm$ 1.8 & 58.1 $\pm$ 1.5 & 10.8 & 48.6  \\
7SW& 251.8 $\pm$ 3.9 & 95.7 $\pm$ 9.1 & 243.2 $\pm$ 0.4 & 70.0 $\pm$ 1.0 & 8.5 & 25.7  \\
&   &  & 313.6 $\pm$ 1.5 & 30.5 $\pm$ 3.6 &  &  \\
7& 249.3 $\pm$ 1.2 & 40.0 $\pm$ 2.9 & 245.1 $\pm$ 0.2 & 65.1 $\pm$ 0.6 & 4.1 & -25.1  \\
&   &  & 318.2 $\pm$ 1.0 & 37.4 $\pm$ 2.3 &  &  \\
8W&   &  & 203.9 $\pm$ 0.8 & 86.5 $\pm$ 2.0 &  &  \\
&  299.5 $\pm$ 1.7 & 44.3 $\pm$ 3.9 & 288.7 $\pm$ 0.1 & 48.2 $\pm$ 0.2 & 10.8 & -3.9  \\
8& 193.0 $\pm$ 3.6 & 98.5 $\pm$ 9.3 & 186.9 $\pm$ 0.6 & 83.3 $\pm$ 1.4 & 6.1 & 15.3  \\
&  302.3 $\pm$ 1.0 & 57.9 $\pm$ 2.4 & 295.9 $\pm$ 0.1 & 54.4 $\pm$ 0.3 & 6.3 & 3.5  \\
9& 167.8 $\pm$ 0.4 & 66.8 $\pm$ 0.9 & 174.3 $\pm$ 0.1 & 60.6 $\pm$ 0.3 & -6.5 & 6.1  \\
&   &  & 276.8 $\pm$ 0.9 & 73.9 $\pm$ 2.2 &  &  \\
9NE& 168.8 $\pm$ 2.4 & 92.7 $\pm$ 6.2 & 149.7 $\pm$ 0.5 & 64.3 $\pm$ 1.0 & 19.1 & 28.4  \\
&  287.9 $\pm$ 5.6 & 73.7 $\pm$ 13.6 & 248.3 $\pm$ 1.4 & 114.9 $\pm$ 3.2 & 39.6 & -41.2  \\
10&  126.1 $\pm$ 5.4 & 68.5 $\pm$ 13.5 & 163.9 $\pm$ 0.4 & 94.4 $\pm$ 1.0 & -37.8 & -25.9  \\
&  267.9 $\pm$ 2.0 & 166.5 $\pm$ 7.1 &  &  &  &   \\
&  269.5 $\pm$ 1.0 & 49.4 $\pm$ 3.3 & 277.0 $\pm$ 0.2 & 58.0 $\pm$ 0.5 & -7.5 & -8.5  \\
10NE& 169.3 $\pm$ 33.6 & 142.8 $\pm$ 44.1 & 176.4 $\pm$ 0.6 & 94.0 $\pm$ 1.3 & -7.1 & 48.8  \\
&  236.8 $\pm$ 1.2 & 53.2 $\pm$ 4.0 & 259.4 $\pm$ 0.3 & 60.8 $\pm$ 0.6 & -22.6 & -7.6  \\
&  306.3 $\pm$ 9.0 & 148.0 $\pm$ 12.0 &  &  &  &  \\
11&  143.1 $\pm$ 3.6 & 116.6 $\pm$ 5.9 & 146.1 $\pm$ 0.1 & 84.4 $\pm$ 0.4 & -2.9 & 32.2  \\
&  260.8 $\pm$ 7.0 & 111.0 $\pm$ 11.3 & 265.8 $\pm$ 0.4 & 62.6 $\pm$ 0.9 & -5.0 & 48.4  \\
12& 142.4 $\pm$ 2.4 & 46.0 $\pm$ 7.7 & 134.2 $\pm$ 0.2 & 63.6 $\pm$ 0.5 & 8.2 & -17.6  \\
&  242.1 $\pm$ 3.6 & 166.9 $\pm$ 7.9 & 220.9 $\pm$ 1.0 & 121.2 $\pm$ 1.9 & 21.3 & 45.7  \\
13&  212.1 $\pm$ 9.3 & 193.0 $\pm$ 18.6 & 211.3 $\pm$ 0.9 & 131.4 $\pm$ 1.4 & 0.8 & 61.7  \\
&  248.9 $\pm$ 0.8 & 54.8 $\pm$ 2.5 & 265.6 $\pm$ 0.1 & 34.5 $\pm$ 0.5 & -16.7 & 20.3  \\
14& 197.2 $\pm$ 0.5 & 68.6 $\pm$ 1.1 & 217.7 $\pm$ 0.1 & 47.9 $\pm$ 0.2 & -20.4 & 20.6  \\
\hline\hline
\end{tabular}
\label{kinematics}
\end{table*}

\subsubsection{The Electron Temperature}

\begin{figure*}
\includegraphics[scale=0.35]{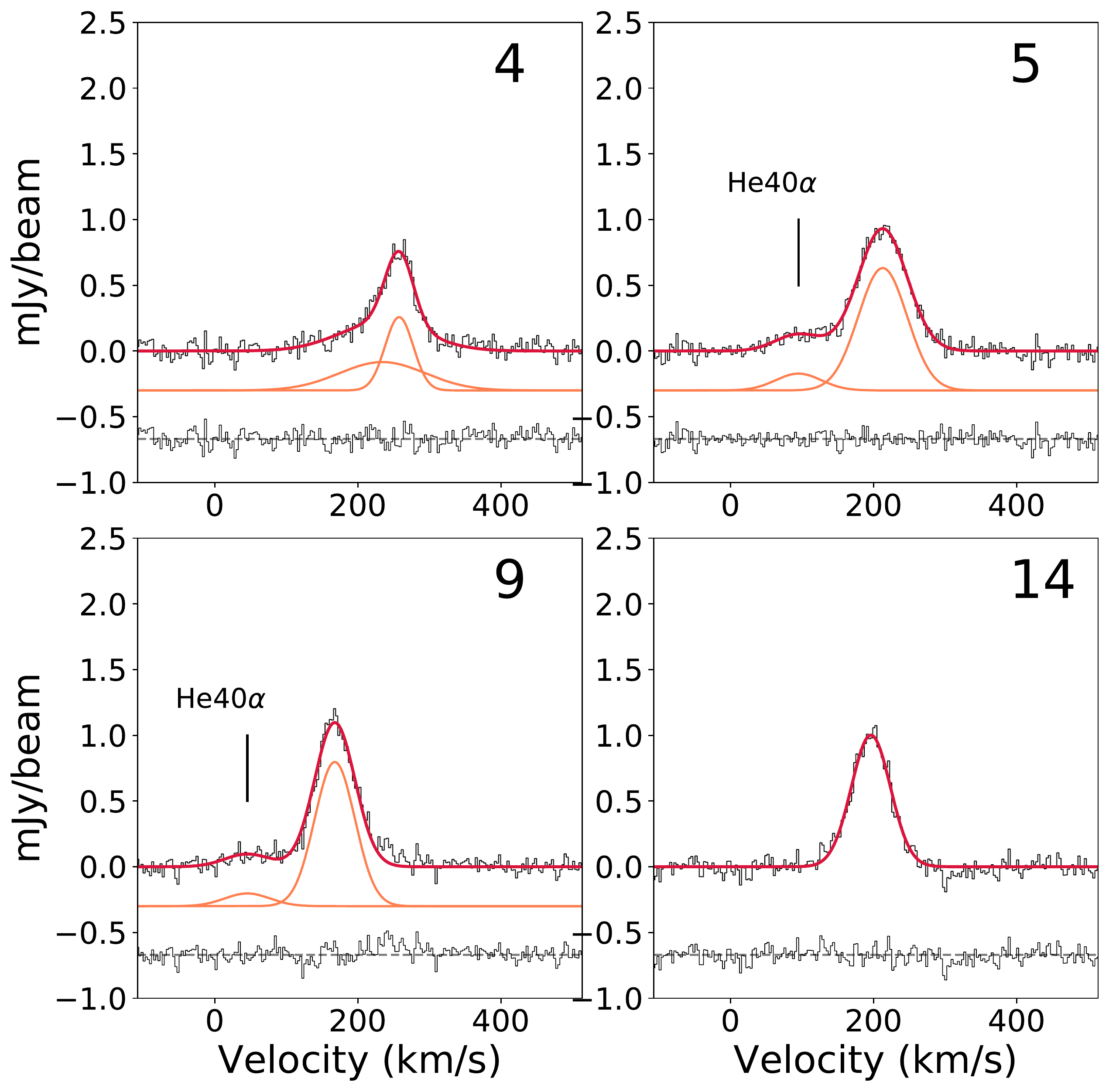}\includegraphics[scale=0.35]{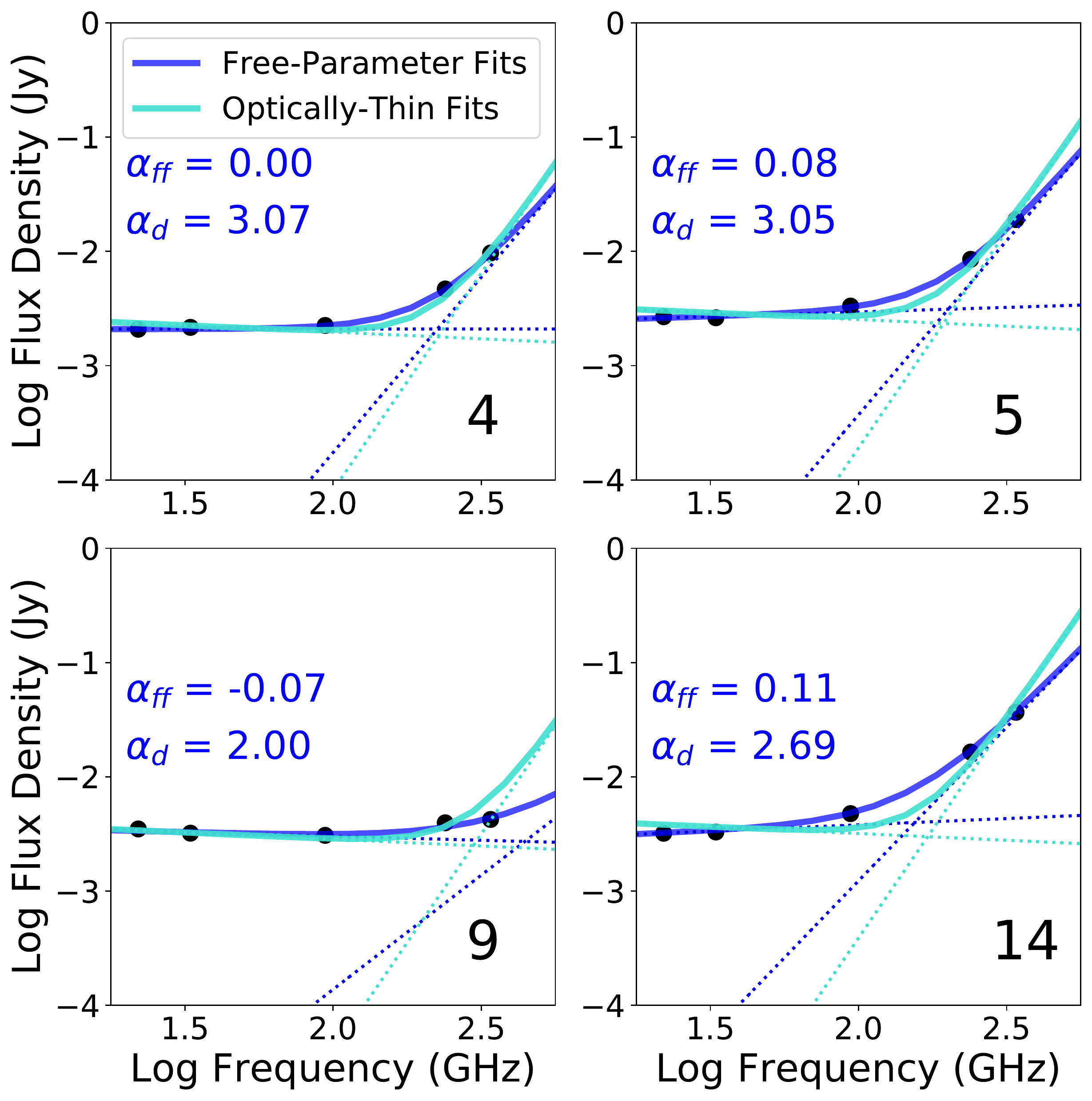}
\caption{{\bf Left:} Line fits to H40 $\alpha$ spectra extracted from the larger ($0\arcsec.3$) apertures used to measure the continuum fluxes for the four isolated sources shown in the second panel of \ref{Fig1}. One and two-component Gaussian fits are conducted, and the best-fit model is overplotted in red. Where two Gaussian components are fit, we show the individual components in orange. The residuals of these fits are plotted below the spectra. {\bf Right:} Fits to the spectral energy distribution composed of continuum measurements at 22 GHz, 33 GHz, 94 GHz, 240 GHz, and 350 GHz. Continuum fluxes are extracted from a $0\arcsec.3$ aperture, and background-subtracted using an annulus with an inner (outer) radius of $0\arcsec.3$ ($0\arcsec.36$). The individual contributions of free-free and dust emission are plotted as dotted lines, while the resulting composite is plotted as a solid curve. The cyan lines show fits that fix the spectral indices of the free-free and dust emission to their optically thin values. The dark blue lines show fits where both spectral indices are allowed to freely vary.}
\label{Fig7}
\end{figure*}

In order to infer the total ionizing flux of the detected sources in the nucleus of NGC 253 from the recombination line emission we must measure a representative electron temperature ($T_e$). Determining $T_e$ requires a measurement of the free-free flux at the same frequency as the H40$\alpha$ recombination line. A complication is that in many sources the 3 mm continuum also has contributions from both synchrotron and dust emission (see, e.g., Figure \ref{Fig2}). To isolate the free-free contribution to the 3 mm continuum flux, we construct five-point spectral energy distributions (SEDs) using our 3 mm (94 GHz) continuum map, 22 GHz and 33 GHz radio maps \citep{Gorski17,Gorski19}, our 350 GHz continuum map from \citetalias{Leroy18}, and an archival continuum image at 240 GHz (project 2012.1.00789.S, PI: K. Nakanishi). We smooth all of these images to a resolution of $0\arcsec.5\times0\arcsec.4$ in order to match the resolution of the 240 GHz map. Because of the decreased resolution, we select only a representative set of sources (4, 5, 9, and 14) that are both reasonably isolated from surrounding sources and exhibit bright H40$\alpha$ emission. We extract continuum fluxes from $r=0\arcsec.3$ (5.1 pc)  apertures centered on these sources and subtract the extended continuum background around these sources using an annulus with an inner radius of $0\arcsec.3$ and an outer radius of $0\arcsec.36$ (as shown in the bottom right panel of Figure \ref{Fig1}). We also smooth our H40$\alpha$ data to match the resolution of the 240 GHz map and extract spectra from a matching $r=0\arcsec.3$ aperture. Fits to these spectra are shown in the left panel of Figure \ref{Fig7}.

As seen in the SEDs shown in the right panel of Figure \ref{Fig7}, all four of the selected sources have negligible synchrotron contribution, which would appear as a steeply decreasing spectral index ($\alpha\sim-0.75$, where F$\propto\nu^{\alpha}$) at the lowest frequencies. Source 9 further has very little associated dust continuum emission, as a strongly increasing spectral index is not seen at the highest frequencies. We fit the SEDs of all four sources for a range of free-free and dust spectral indices. Given the limited number of constraints available from the five independent frequencies at which we have data, we do not include possible contributions from Anomalous Microwave Emission (AME) in these fits. AME has been observed in extragalactic sources (though not in galaxy nuclei) and peaks at frequencies between 20 and 40 GHz \citep{Hensley15}. Checking for the presence of AME will require observations of NGC 253 at additional radio frequencies, which would enable more detailed SED fitting of all of the clusters. 

To fit the SEDs, we first consider a case in which both the free-free emission and the dust emission are optically-thin. For optically-thin free-free emission at radio and microwave wavelengths, we expect a power law with a slightly negative spectral index of $\alpha_{ff}$ = -0.118 \citep{Draine}. This is consistent with the spectral index adopted by \cite{RodriguezRico06} in their analysis of 7 mm data of NGC 253, though it is slightly shallower than the spectral index adopted by \cite{Bendo15} in fits to ALMA 3 mm observations on 35 pc scales. \cite{Bendo15} use a value of $\alpha_{ff}$=-0.17 from \cite{Scoville13}, however this is more appropriate at sub-millimeter wavelengths, rather than the predominately longer wavelengths at which we primarily detect free-free emission. For optically-thin dust emission at radio and millimeter wavelengths we expect a power law with a spectral index $\alpha_{d}$. This spectral index is typically expressed as 2$+\beta$, where $\beta$ is the dust emissivity coefficient, and is expected to be between 1.5 and 2. We fix the spectral index to be 3.8 (with $\beta$ assumed to be comparable to measurements of $\beta$=1.8 measured from blackbody fits to $\sim$ 1 pc resolution observations of the Galactic center embedded cluster Sgr B2; \citealt{Schmiedeke16}). Restricting the spectral indices to these values, we fit for the intercepts of both power-law distributions, with the results shown in the right panel of Figure \ref{Fig7}. Our fits assuming optically-thin emission result in free-free contributions to the 3 mm flux ranging from 67-93\% (see Table \ref{T_es}).

We also conduct SED fitting which allows both spectral indices to vary freely. In this case, both the free-free spectral index and the dust spectral index are allowed to be optically-thick. Note that in the optically-thick limit, the dust spectral index is expected to approach a value of 2. Optically-thick free-free emission has a positive, increasing spectral index (up to $\alpha$ = 2) at lower frequencies. This is seen in the super star clusters in NGC 5253 \citep{Meier02,Turner00,Turner98} as well as in the Sgr B2 protocluster at the center of the Milky Way, which has a steeply rising spectral index at frequencies up to 22 GHz, turning over near 33 GHz \citep{Protheroe08}, and for which individual hypercompact HII regions have rising spectral indices up to 60-100 GHz \citep{Zhao11}. The best-fit free-free spectral indices range from -0.1 in source 9 (consistent with optically-thin free-free emission) to 0.1 in source 14. The best-fit dust spectral indices in sources 4 and 9 could be consistent with optically-thin emission, though the spectral index is poorly constrained in source 9. Sources 5 and 14 have spectral indices $<$3.5, indicating somewhat optically-thick dust emission. This is consistent with these clusters being the strongest sources of both 350 GHz dust continuum emission and HCN 4-3 emission \citep{Leroy18}. The free-free contributions to the 3 mm flux corresponding to fits for these four sources range from 79-99\%. For comparison, free-free fractions at 100 GHz for the Sgr B2 protocluster from \cite{Schmiedeke16} range from 0.3 (for the extremely embedded North sub-cluster) to 0.7 (for the more evolved Main subcluster).  

The free-parameter fits to the SEDs of sources 5 and 14 are not consistent with optically-thin dust emission. As can be seen in Figure \ref{Fig7}, the optically-thin fits for these sources underestimate the 94 GHz continuum flux and overestimate the 350 GHz continuum flux. We thus prefer the results of the free-parameter fits for these sources, and for consistency, we adopt the results of the free-parameter fits for all 4 sources in our remaining analysis.
We first use the estimated free-free contributions to the 3 mm continuum flux to derive electron temperatures for sources 4, 5, 9, and 14. To determine the electron temperature $T_e$ we use the formula derived in Appendix A of \cite{Emig20}, based on \cite{Draine} and \cite{GS02}:

\begin{equation}
\begin{split}
T_e =  10^4\; \mathrm{K} \biggl[ b_{n+1}\; \left(\frac{1}{1+y^+}\right) \left(\frac{R_{LC}}{31.31\; \mathrm{km s}^{-1} } \right)^{-1}\times\\
 \left(\frac{\nu}{100\; \mathrm{GHz}} \right)^{1.118} \biggr]^{0.85},
\end{split}
\label{Teq}
\end{equation}

where $b_{n+1}$ is the non-LTE departure coefficient, $y^+$ is the measured abundance ratio of He+ to H+, by number ($N_{He^+}/N_{H^+}$), $R_{LC}$ is the ratio of the line-to-continuum emission, and $\nu$ is the frequency of the recombination line in GHz.

$b_{n+1}$ is determined via interpolation from values given in \cite{SH95}. The departure coefficients vary with electron density, and we use coefficients corresponding to densities of $n_e=10^4$ cm$^{-3}$ and $n_e=10^5$ cm$^{-3}$ to determine a range of $T_e$ values. Our choice of electron densities is primarily based on prior measurements of $n_e$ for HII regions in NGC 253, which find $n_e\ge10^4$ cm$^{-3}$ \citep{Ulvestad97,Mohan05}. In particular, \cite{Mohan05} found that only 10\% of the H40$\alpha$ emission can be modeled with a $n_e=10^4$ cm$^{-3}$ component, and suggested that the presence of higher-density ionized gas is required to fit this emission. This is also consistent with modeling of the radio and millimeter continuum emission toward the Sgr B2 protocluster, which is composed of numerous (ultra)compact HII regions with $n_e$ ranging from $n_e=10^4$ cm$^{-3}$ to $n_e=10^6$ cm$^{-3}$ \citep{Schmiedeke16,SM17}. 

$y^+$ is  is related to the mass weighted abundances $Y^+$ reported in Table \ref{params} by the equation:

\begin{equation}
Y^+ = \frac{y^+ M_{He}}{y^+ M_{He} + M_H }.
\end{equation}

The ratio of the line-to-continuum emission $R_{LC}$ is given by:

\begin{equation}
R_{LC}= \frac{\int T_L \Delta v}{f_{ff}T_C},
\end{equation}

where $T_C$ is the brightness temperature of the 3 mm continuum, $T_L$ is the brightness temperature of the recombination line, $f_{ff}$ is the fraction of the 3 mm continuum that comes from free-free emission, and $\Delta v$ is the FWHM line width of the recombination line in \kms.

The values of $T_e$ derived from this equation range from 6400 to 11000 K, with a mean value of $\langle T_e \rangle$ = 8000 K. We report these values in Table \ref{T_es}. For comparison, we also report the electron temperatures that would be determined by adopting the free-free fraction from the optically-thin SED fits. These are slightly lower, but are generally consistent given the uncertainties in $T_e$ due to the range of $n_e$ values we adopt for determining the departure coefficient.  The electron temperatures we measure are higher than the values measured in NGC 253 using lower-resolution data \citep[3700-4500 K;][]{Bendo15}. One contributing source of this discrepancy is the higher electron density we have adopted to determine the departure coefficient, as \cite{Bendo15} assume $n_e=10^3$ cm$^{-3}$. The electron temperatures we infer are quite consistent with values measured for \hii regions in the Milky Way's central 200 pc \citep[typically 6000-8000 K, with values $>10,000$ K in individual hypercompact HII regions;][]{Goss85,dePree96,Lang01}. These values are representative of the electron temperature gradient observed in the Milky Way disk, which reaches a minimum at the Galactic center that corresponds with a maximum in the radial metallicity gradient \citep{Quireza06}. However, low values are also seen in some other massive star forming regions in the Milky Way disk \citep[e.g., 8500 K in W51;][]{Ginsburg15}. 

\subsubsection{Revised Cluster Ionizing Fluxes}

We estimate the flux of Lyman continuum photons ($Q_c$) in s$^{-1}$ for each embedded cluster using the following expression, which is valid at high frequencies where the free-free continuum opacity $\tau_{ff} <<1$ \citep{Rubin68,Condon92,Ulvestad97}: 

\begin{equation}
\begin{split}
    Q_c =  7.5\times10^{49} \left(\frac{D}{\mathrm{Mpc}}\right)^2 \left(\frac{T_e}{10^4\; \mathrm{K}}\right)^{-0.45} \times \\
    \left(\frac{\nu}{\mathrm{GHz}}\right)^{0.1} \left(\frac{S_{\nu,ff}}{\mathrm{mJy}}\right).
\end{split}
\end{equation}

Here $\nu$ is the representative continuum frequency in GHz, $T_e$ is the electron temperature (for sources for which we do not have a measurement of $T_e$, we adopt $\langle T_e \rangle$ = 8000 K), $D$ is the distance to the source in Mpc, and $S_{\nu,ff}$ is the free-free flux at 3 mm. For sources 4, 5, 9, and 14 we have estimates of the fraction of the 3 mm continuum flux that is due to free-free emission from the SED fitting. For the remaining sources we use the adopted $\langle T_e \rangle$ to invert Equation \ref{Teq} in order to determine a mean line-to-continuum ratio $\langle R_{LC}\rangle$. For $\langle T_e\rangle = 8000^{+300}_{-1500}$ K (covering the range of temperature variations for the four sources we measure), we derive $\langle R_{LC}\rangle = 28^{+5}_{-2}$ \kms. We then apply this ratio to the measured H40$\alpha$ line fluxes to derive the corresponding free-free flux at 3 mm:

\begin{equation}
S_{\nu,ff} = \frac{\int S_{\mathrm{H}40\alpha} \Delta v}{\langle R_{LC}\rangle}.
\end{equation}

The resulting $Q_c$ values are presented in Table \ref{stellar}. Typical uncertainties on these values due to the uncertainty in the adopted $T_e$ are 0.1-0.2 dex. We compare these to the \citetalias{Leroy18} estimates of $Q_c$ from the 33 GHz continuum, which assumed that the 33 GHz fluxes are entirely due to free-free emission. For the majority of the detected sources, the ionizing fluxes we infer from the H40$\alpha$ line are consistent with the fluxes determined by \citetalias{Leroy18}, within the uncertainties deriving from our temperature measurements.  However, there are several sources where the Lyman continuum fluxes we derive are substantially smaller than the \citetalias{Leroy18} values. The largest difference is seen for source 12, for which we estimate a $Q_c$ that is nearly an order of magnitude less than estimated by \citetalias{Leroy18}. This source was flagged by \citetalias{Leroy18} as having an abnormally high free-free contribution to the 350 GHz continuum (71\%), based on the assumption that the 33 GHz flux is entirely due to free-free emission. Given the relatively small amount of free-free emission at 3 mm that is implied by the H40$\alpha$ fluxes, combined with the negative spectral index seen for this source in Figure \ref{Fig2}, our measurements indicate that the 33 GHz continuum for this source has a substantial synchrotron component. The same appears to be true for sources 1, 6, 10 and 11, though to a significantly lesser extent than in source 12. These sources have smaller discrepancies, being lower than the 33 GHz continuum-derived $Q_c$ by 0.3-0.5 dex. Sources 6 and 11 were also noted by \citetalias{Leroy18} as having apparently large free-free contributions to the 350 GHz continuum  flux (63 and 41\%, respectively). In contrast, source 10 had only a moderate estimated free-free contribution (11\%), and source 1 had one of the lowest estimated free-free contributions to the 350 GHz continuum, only 3\%. We further discuss these results and possible sources of uncertainties in both methods of deriving $Q_c$ in Section \ref{SF}.

\begin{table}[t]
\caption{Electron Temperatures}
\centering
\begin{tabular}{lllll}
\hline\hline
{Source} & 4 & 5 & 9 & 14 \\
\hline
$\alpha_{dust}$\tablenotemark{a} & 3.8 & 3.8 & 3.8 & 3.8 \\
$\alpha_{ff}$\tablenotemark{a} & -0.118 & -0.118 & -0.118 & -0.118 \\
$f_{ff}$ & 0.87 & 0.78 & 0.93 & 0.68  \\
$T_e$\tablenotemark{b} & 6600 $\pm$ 700 & 6600 $\pm$ 700  & 7400 $\pm$ 800  & 9100 $\pm$ 900 \\
\hline
$\alpha_{dust}$ & 3.1 $\pm$ 0.9 & 3.0 $\pm$ 0.5 & $<$ 3.7 & 2.6 $\pm$ 0.3 \\
$\alpha_{ff}$ &0.0 $\pm$ 0.1 & 0.1 $\pm$ 0.1 & -0.1 $\pm$ 0.1 & 0.1 $\pm$ 0.1 \\
$f_{ff}$ & 0.95 & 0.89 & 1.01 & 0.76  \\
$T_e$\tablenotemark{b} & 7100 $\pm$ 700 & 7400 $\pm$ 800  & 7900 $\pm$ 800  & 10000 $\pm$ 1000 \\
\hline\hline
\end{tabular}
\tablenotetext{a}{Spectral indices fixed to optically-thin values}
\tablenotetext{b}{$T_e$ ranges determined for $n_e=10^4$ cm$^{-3}$ and $n_e=10^5$ cm$^{-3}$}
\label{T_es}
\end{table}

\section{Discussion}
\label{dis}

Below, we include additional discussion of several of the results from these new ALMA observations including the detection of broad-linewidth hydrogen recombination line emission from a number of the embedded clusters, and the update of embedded stellar cluster properties compared to the \citetalias{Leroy18} values derived from the radio continuum emission. 

\subsection{Kinematic comparison with HCN 4\texorpdfstring{$-$}{-}3}

\begin{figure*}
\includegraphics[scale=0.21]{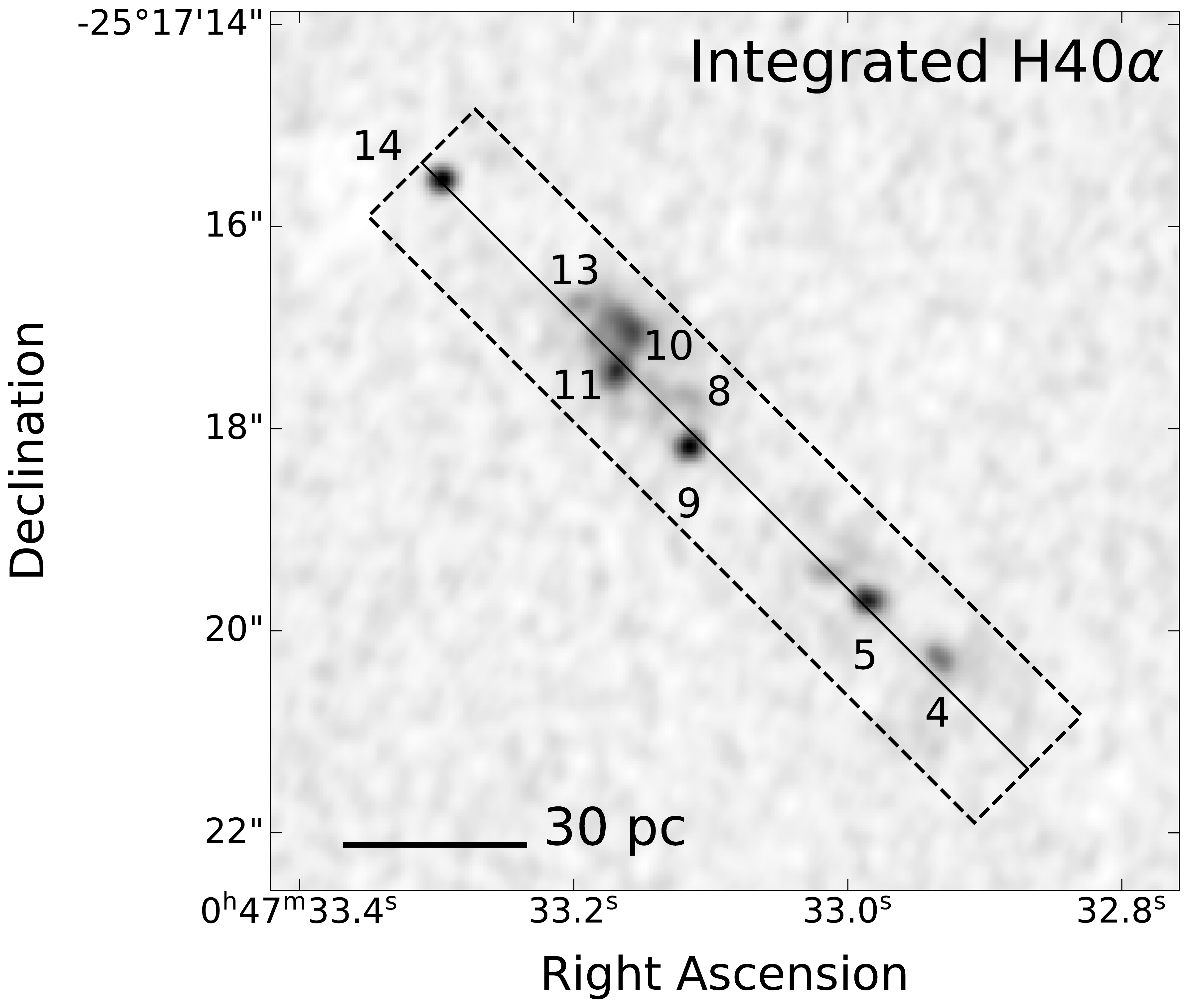}\includegraphics[scale=0.41]{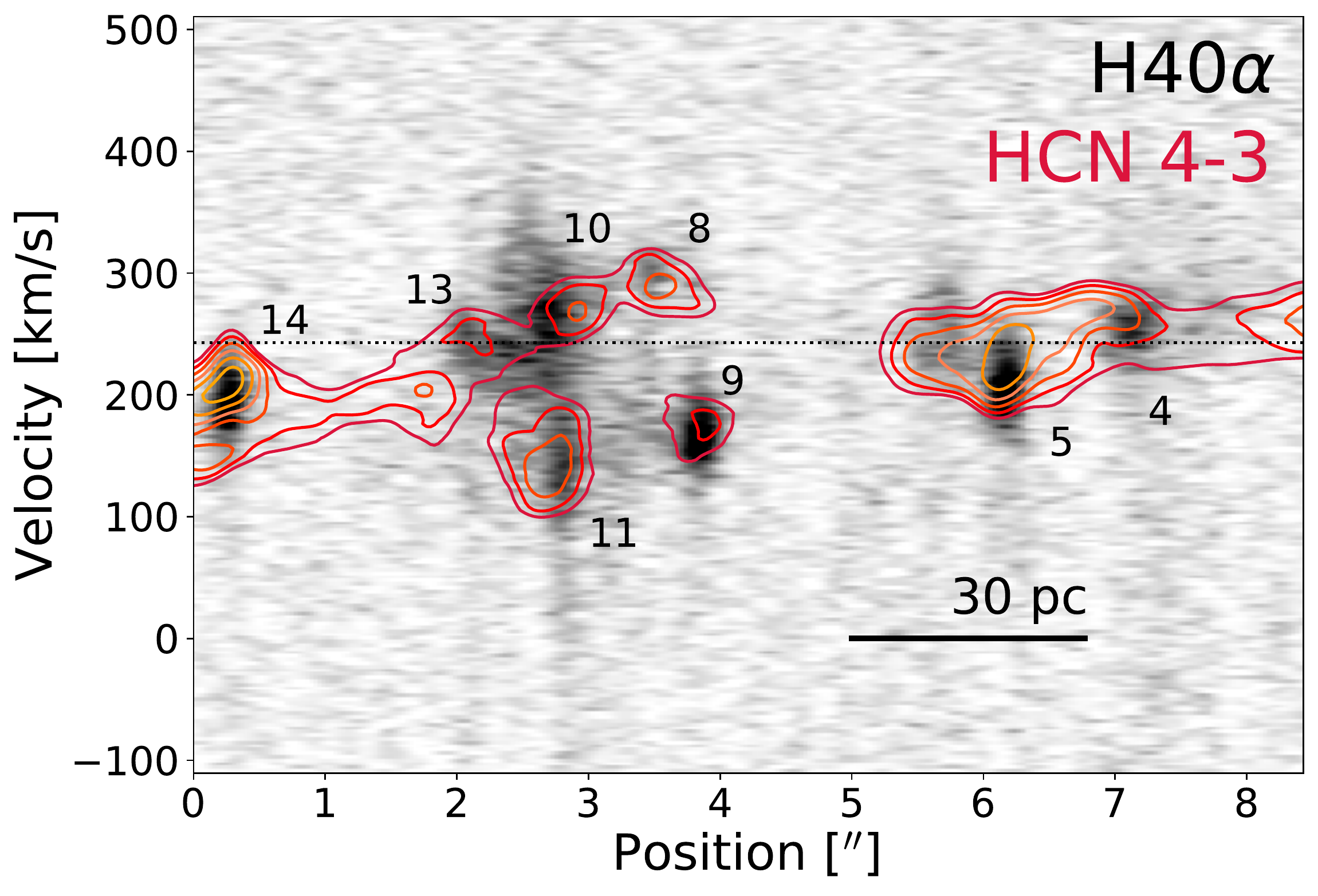}
\caption{{\bf Left:} Map of integrated H$\alpha$ emission showing the orientation and width of the slit used to extract a position-velocity diagram. {\bf Right:} Position-velocity diagram of H40$\alpha$ emission (greyscale) compared to HCN 4-3 emission (red contours) along the same slit. The horizontal dotted line shows the rest-frame velocity of NGC 253. Emission at large positive-velocities is seen between source 10 and 13 in H40$\alpha$ but not in HCN 4-3. While there also appears to be large negative-velocity emission extending away from many of the sources (e.g., 4, 5, 9, 11, 13) this is contaminated by the presence of He40$\alpha$ emission.}
\label{Fig8}
\end{figure*}

To facilitate comparison between the kinematics of the ionized gas traced by H40$\alpha$ and the dense molecular gas traced by HCN 4$-$3, we construct a position-velocity (PV) diagram of the central emission along the plane of NGC 253 in both lines (Figure \ref{Fig8}). Examining the PV diagram, we see that emission in H40$\alpha$ extends beyond HCN 4$-$3 for many sources, consistent with our analysis of the spectra, which found that the H40$\alpha$ emission is broader than the HCN 4$-$3 emission in the majority (12/18) of the sources. We note that the lower-velocity extent of the H40$\alpha$ emission seen in Figure \ref{Fig8} does have a somewhat ambiguous interpretation, as this can include a contribution from He40$\alpha$ (which is offset by -122 \kms) as was identified in sources 5, 9, and 14. However, this contamination is not present at larger velocities, and we see a clear excess of high-velocity emission toward sources 10, 10NE, and 6. Toward source 10NE in particular, there is a near complete absence of corresponding HCN emission at velocities $\gtrsim 270$ \kms. The PV diagram also shows the presence of systematic velocity offsets in sources 4, 5, and 14, where the H40$\alpha$ emission appears to be shifted to lower velocities by $\sim$ 20 \kms\, compared to the HCN 4$-$3 emission. As described in detail in \cite{Levy21}, these three clusters all show signatures of outflows in higher-resolution data of dense molecular tracers, including HCN 4$-$3. 

As mentioned in Section \ref{recomb}, we find broad ($\Delta v_{FWHM}> 100$ \kms) H40$\alpha$ line emission toward 7 of the sources studied here: 4, 6, 10, 10NE, 11, 12, and 13.  A complication in interpreting these apparently broad linewidths, illustrated in the PV diagram, is that several of these sources are on lines of sight that intersect multiple distinct kinematic components. If these components are not well resolved, these can lead to an apparent broadening of the observed line profiles \citep{Leroy15}. These kinematic components may be spatially separated, for example corresponding to the near and far sides of an `x2' family of orbits \citep{Binney91,Sormani15}, analogous to the `twisted ring' or `orbital streams' seen in the Milky Way center \citep{Molinari11,Kruijssen15,Henshaw16}. Such nuclear rings, with radii of a few hundred parsecs, are ubiquitously observed at the centre of many barred galaxies \citep[e.g., ][]{Comeron10} and their presence is explained by models of gas flow in barred galaxies \citep{KK15,Torrey17,Sormani20}. In NGC 253, sources 8 and 13 are characteristic of the high-velocity gas, with their strongest emission at velocities of 250-300 \kms, while sources 5 and 9 are characteristic of the low-velocity gas, with their strongest emission peaking between $\sim$150-200 \kms. As both velocity components may be present toward many of the embedded clusters, we have made sure to fit complex line profiles with both a low and high-velocity component, so that this kinematic complexity is not interpreted as an exaggerated line width. We find that the H40$\alpha$ line widths toward these sources, even when fitting multiple components to the line profiles, are significantly broader than observed toward the other clusters (which have $\Delta v_{FWHM}< 100$ \kms).

\subsection{The Nature of the Broad-Linewidth H40\texorpdfstring{$\alpha$}{-alpha} Emission}

We suggest that the large velocity extent of the H40$\alpha$ line seen in sources 4, 6, 10, 10NE, 11, 12, and 13 is evidence that the embedded clusters in the nuclear starburst of NGC 253 are contributing to driving the hot ionized wind component of the multiphase outflow. The observed widths of the H40$\alpha$ line in these sources ( $\Delta v_{FWHM} \sim 100-200$ \kms; $\sigma \sim40-90$ \kms) are substantially larger than the virial linewidths for these clusters ($\sigma \sim$ 20 \kms). 
The H40$\alpha$ is also $\sim50$ \kms broader than the HCN 4$-$3 emission at a comparable velocity, implying that the ionized gas is substantially less bound to the cluster than the  dense \citep[$\gtrsim10^3$ cm$^{-3}$;][]{Kauffmann17} molecular gas, which may still be fueling star formation in the clusters. 
We further detect a broad-linewidth component in sources 10 and 10 NE at large positive velocities (270-300 \kms) that has no clear counterpart in HCN 4$-$3. If the observed broad linewidths are tracing the hot wind, this component could correspond to compact emission associated with the northern extension of the inner superwind, which has been previously undetected due to obscuration from the foreground disk of the galaxy \citep{Sharp10,Westmoquette11}. This would represent the counterpart to the southern extension seen in X-ray and ionized gas tracers with velocities of 100$\pm$50 \kms \citep{Sharp10,Westmoquette11,Pietsch00}. While these seven sources represent the clearest examples of broad linewidth H40$\alpha$ emission, several other sources (5, 7SW, 8, 9NE, and 14) also have H40$\alpha$ emission that is 10-30 \kms broader than the HCN 4$-3$ emission.

We note that the high-velocity, broad-linewidth emission detected toward sources 10, 10NE, 11 and 12 originates in a 1$''$.2 (20 pc) region that overlaps with the apparent kinematic center of NGC 253 \citep{MullerSanchez10}. Some of the large velocity dispersion observed in the H40$\alpha$ line toward these sources could then be due at least in part to the range of orbital velocities expected for sources and/or extended ionized gas lying within 5-10 parsecs of a supermassive black hole, similar to the velocity spread ($\pm 200$ \kms) seen for molecular gas in the circumnuclear disk of the Milky Way \citep[e.g., ][]{Chris05}. However, a complication of this interpretation is that the orientation of the velocity gradient of this gas is perpendicular to the geometry of the overall nuclear disk. While this is not impossible, we view it as more likely that a velocity gradient perpendicular to the disk would be associated with an outflow. 

If the high-velocity emission in the vicinity of sources 10, 10NE, 11 and 12 were to originate from the central few parsecs, this outflow is unlikely to be driven by the black hole itself, as there is no indication of an active black hole, which would be expected to emit simultaneously at X-ray, radio, and IR wavelengths \citep{Engelbracht98,Brunthaler09,FO09,MullerSanchez10}. A weak  AGN is also unlikely to be responsible for ionizing this gas, given that observations of submillimeter recombination lines around the luminous Type 2 Seyfert NGC 1068 have failed to detect emission \citep{Izumi16}. The AGN in starburst galaxy NGC 4945 has also been shown not to dominate the ionization of the nuclear region \citep{Spoon00,Marconi00,Emig20}. Models further predict that in the presence of an AGN, recombination lines from He$^{++}$ would be stronger than the hydrogen recombination lines \citep{Scoville13}, and we do not detect any emission from the HeII (64) $\alpha$ line at 98.0795642 GHz.

\subsection{Clustered Star Formation in NGC 253}
\label{SF}

\begin{table}[t]
\caption{Stellar Properties of Clusters}
\centering
\begin{tabular}{lll|ll|l}
\hline\hline
   & \multicolumn{2}{c}{$log_{10}$ Q$_{tot}$ }  & \multicolumn{2}{c}{$log_{10}\, M_*$ } & $log_{10}\, M_{gas}$ \\
   & \multicolumn{2}{c}{(s$^{-1}$) }  & \multicolumn{2}{c}{($M_\odot$)} & ($M_\odot$) \\
ID &  H40 $\alpha $ & 33 GHz\tablenotemark{a} &  H40 $\alpha $ & 33 GHz\tablenotemark{a} & HCN 4$-$3\tablenotemark{a} \\
\hline
1& 50.6$_{-0.2}^{+0.2}$ & 50.9 & 4.0$_{-0.2}^{+0.2}$ & 4.3 & 4.9 \\
2& 50.8$_{-0.2}^{+0.2}$ & 50.9 & 4.2$_{-0.2}^{+0.2}$ & 4.3 & 4.7 \\
3& 50.8$_{-0.2}^{+0.2}$ & 50.7 & 4.2$_{-0.2}^{+0.2}$ & 4.1 & 5.1 \\
4& 51.7$_{-0.1}^{+0.1}$ & 51.6 & 5.1$_{-0.1}^{+0.1}$ & 5.0 & 5.1 \\
5& 51.9$_{-0.1}^{+0.1}$ & 52.0 & 5.3$_{-0.1}^{+0.0}$ & 5.4 & 5.3 \\
6& 51.6$_{-0.2}^{+0.2}$ & 51.9 & 5.0$_{-0.2}^{+0.2}$ & 5.3 & 3.6 \\
7SW& 51.3$_{-0.2}^{+0.2}$ & & 4.7$_{-0.2}^{+0.2}$ & & \\
7& 50.9$_{-0.2}^{+0.2}$ & 51.1 & 4.3$_{-0.1}^{+0.3}$ & 4.5 & 4.5 \\
8W& 51.0$_{-0.2}^{+0.2}$ & & 4.4$_{-0.2}^{+0.2}$ & & \\
8& 51.6$_{-0.2}^{+0.2}$ & 51.4 & 5.0$_{-0.1}^{+0.2}$ & 4.8 & 5.2 \\
9& 51.9$_{-0.1}^{+0.1}$ & 52.1 & 5.3$_{-0.1}^{+0.0}$ & 5.5 & 4.7 \\
9NE& 51.5$_{-0.2}^{+0.2}$ & & 4.9$_{-0.2}^{+0.2}$ & & \\
10& 51.4$_{-0.2}^{+0.2}$ & 51.9 & 4.8$_{-0.2}^{+0.2}$ & 5.3 & 5.2 \\
10NE& 51.5$_{-0.2}^{+0.2}$ & & 4.9$_{-0.1}^{+0.2}$ & & \\
11& 51.9$_{-0.2}^{+0.2}$ & 52.2 & 5.3$_{-0.2}^{+0.2}$ & 5.6 & 4.5 \\
12& 51.7$_{-0.2}^{+0.2}$ & 52.6 & 5.1$_{-0.2}^{+0.2}$ & 6.0 & 4.1 \\
13& 51.7$_{-0.2}^{+0.2}$ & 51.4 & 5.1$_{-0.2}^{+0.2}$ & 4.8 & 5.2 \\
14& 51.9$_{-0.1}^{+0.1}$ & 52.1 & 5.3$_{-0.0}^{+0.1}$ & 5.5 & 5.7 \\
\hline
Total & 52.8 & 53.1 & 6.2 & 6.5 & 6.2 \\
\hline\hline
\end{tabular}
\tablenotetext{a}{Values from \cite{Leroy18}}
\label{stellar}
\end{table}

\citetalias{Leroy18} identified a population of 14 compact sources of millimeter dust and spectral line emission toward the nucleus of NGC 253, which they interpreted as embedded clusters, based on the close association of the dust continuum emission at 350 GHz with radio continuum emission at 33 GHz. Only one of these sources was previously known (Source 5) from HST observations, where it was identified as a highly obscured super star cluster \citep{Watson96,Kornei09}. Assuming that the 33 GHz continuum is entirely dominated by free-free continuum, \citetalias{Leroy18} estimated the ionizing fluxes of these clusters and their corresponding stellar masses. 

Using observations of the H40$\alpha$ recombination line we have reassessed the properties of the embedded clusters, testing the assumption that the 33 GHz radio continuum is dominated by thermal free-free emission. Using the recombination line fluxes and new measurements of the electron temperature $T_e$ for four of these sources, we have made independent estimates of $Q_c$ for the 14 sources identified by \citetalias{Leroy18}, as well as 4 newly-identified sources. We find that sources 1, 6, 10, 11, and 12 all have a lower Lyman continuum flux than estimated by \citetalias{Leroy18}. In most cases, this is likely because the 33 GHz continuum contains an additional contribution from synchrotron emission (see Figure \ref{Fig2}), as was postulated by \citetalias{Leroy18} for a number of these sources, based on the apparently large amount of free-free flux that would be present at 350 GHz, if the 33 GHz continuum flux were entirely due to free-free emission. However, we also find that an apparently large contribution of free-free emission as extrapolated from the 33 GHz flux to the flux at 350 GHz is not necessarily an indication of substantial synchrotron contamination. Some sources likely do have a significant free-free contribution at 350 GHz, as they do not show negative spectral indices in Figure \ref{Fig2}. For example, the SED of source 9 (Figure \ref{Fig7}) shows extremely weak dust emission compared to sources 4, 5, and 14, consistent with the estimate by \citetalias{Leroy18} that the free-free contribution at 350 GHz would be 32\%. All together, we find a slightly lower total ionizing flux contribution from the embedded clusters (log[$Q_{tot}$] = 52.8) compared to \citetalias{Leroy18}, who estimated log[$Q_{tot}$] = 53.1. That this difference is not larger is due primarily to the contribution from the four sources not catalogued by \citetalias{Leroy18}. 

We note however that significant uncertainties remain in the determination of the Lyman continuum flux. Firstly, all $Q_c$ values, both those derived from 33 GHz and from recombination lines, should be treated as lower limits, as any dust in the HII region will directly absorb some fraction of the Lyman continuum photons, reducing the number of ionizations and leading to an underestimate of $Q_c$. We also note that while recombination line-derived $Q_c$ values which are less than \citetalias{Leroy18} can be well explained as a result of synchrotron emission, we also find a $Q_c$ for source 13 that is significantly larger than in \citetalias{Leroy18}. This is more difficult to explain, but could be the result of somewhat optically-thick 33 GHz emission, or perhaps a significantly lower electron temperature, though our uncertainties cover a wide range of $T_e$ from $\sim$6000-11000 K. Ultimately, reducing this uncertainty in the $T_e$ measurements for the embedded clusters is essential for future work. 

There are two main sources of uncertainty in the $T_e$ measurements. First, the electron temperature depends sensitively on the assumed electron density $n_e$ which is used to determine the departure coefficients $b_n$. Parsec-scale observations of additional recombination lines would help with this by allowing us to model the electron density. Sub-parsec imaging of the thermal gas (either from radio/mm continuum or recombination lines) would also be useful to better constrain source sizes, which would give an independent constraint on the electron densities. Currently, sub-parsec imaging has only been completed for the dust continuum associated with these clusters \citep{Levy21}. Second, the measurement of the electron temperatures requires full SED modeling to make a good estimate of the amount of free-free emission contributing to the total continuum emission. Accurately determining the free-free continuum in all of the embedded cluster sources will require parsec-resolution measurements of the continuum emission at additional submillimeter wavelengths. Lacking this, we cannot accurately determine the amount of free-free emission at radio and millimeter wavelengths. 

 Using our $Q_c$ values, we determine a corresponding stellar mass for the embedded clusters, following the same method  as \citetalias{Leroy18} and calculating a mass based on Starburst99 calculations for a ZAMS  stellar population \citep{Leitherer99}. The stellar mass $M_*$ relates to the rate of production of ionizing Lyman continuum photons ($Q_c$) as:

\begin{equation}
    M_* \sim \frac{Q_c}{4\times10^{46}} \mathrm{M}_\odot.
\end{equation}

The masses we derive are sensitive to the assumption of the age of the stellar population. Cluster masses derived for NGC 4945 \citep{Emig20}, for which the estimated age of the starburst is 5 Myr, are substantially more massive as an older stellar population produces fewer ionizing
photons per unit mass. Adopting a burst age of 5 Myr for NGC 253 would increase the masses derived here by about an order of magnitude. However, our observations continue to support the adoption of a ZAMS population, as originally argued by \citetalias{Leroy18}. Our detection of the He recombination line emission favors a young cluster age, as the H-ionizing radiation from massive stellar populations at high metallicities decreases steeply after $\sim$3 Myr \citep[e.g.,][]{Levesque13} and the He-ionizing radiation decreases even more steeply \citep[e.g.,][]{Levesque12}. Additionally, we can also make a direct comparison with NGC 4945 to determine a relative age for the two bursts. As seen in \cite{Emig20}, there is substantially more nonthermal emission (due to supernova remnants) intermixed with the cluster emission. The clusters in NGC 4945 also have smaller gas fractions (typically an order of magnitude less than the stellar mass). Together, these differences support the assumption that the cluster ages in NGC 253 are significantly younger.

As expected, the stellar masses of the embedded clusters that we infer from the $Q_c$ values are also generally lower than the masses estimated by \citetalias{Leroy18}. Once again, the largest difference is for source 12: the stellar mass we infer for this source is lower by a factor of 8, making it no longer the most massive source. The most massive embedded clusters are now sources 9, 11, and 14 (each with an inferred stellar mass of 2$\times10^5$ M$_\odot$). The new sources (7SW, 8W, 9NE, and 10NE) have inferred stellar masses of 3-8$\times10^4$ M$_\odot$. While low, these masses are still larger than found for embedded clusters 1, 2, 3, and 7, which have inferred stellar masses of 1-2$\times10^4$ M$_\odot$. 

Overall, we find a minimum total ionizing flux from the clusters of $Q_c=6.3\times10^{52}$ s$^{-1}$ or 20\% of the total ionizing flux of the nuclear starburst as originally estimated by \cite{Bendo15}. However, \cite{Bendo15} make a number of different assumptions in their calculation of $Q_c$ that make a direct comparison invalid. They assume a slightly smaller distance (3.44 Mpc vs. 3.5 Mpc), a lower electron temperature (3900 K vs. 8000 K) and electron density ($10^3$ cm$^{-3}$ vs. $10^4$ cm$^{-3}$), and their expression for $Q_c$ includes a frequency dependence of $\nu^{0.17}$ instead of $\nu^{0.1}$. If the calculation of $Q_c$ in \cite{Bendo15} via their Equation 3 instead adopted the values from our analysis, then their estimated $Q_c$ would be smaller by a factor of 2.4. We then find that the clusters contribute at least 50\% of the total ionizing flux of the starburst. This result confirms the finding of \citetalias{Leroy18} that the embedded massive clusters represent an important mode of the overall starburst, which may even be the dominant mode, depending on the extent to which the $Q_c$ we measure is affected by dust in the embedded clusters, or whether there is a non-zero Lyman continuum escape fraction. This result is also consistent with findings that clustered star formation is a preferred mode in starburst environments in the Milky Way center, where a high cluster formation efficiency ($\sim 40\%$) is measured in the densest part of the Sgr B2 protocluster \citep{Ginsburg18}.

\subsection{Evolution of the Starburst}
 
Characterizing the age and evolution of the central starburst in NGC 253 is important both for inferring the current overall stellar content of the clusters and for determining how the evolution of the starburst drives the observed outflows emanating from the nucleus. The candidate protoclusters in the NGC 253 nucleus appear to still be embedded in or associated with their natal molecular clouds. \citetalias{Leroy18} find that 7 out of the 14 embedded clusters they identify have higher gas masses than stellar masses. From our new estimate of the total stellar masses of the embedded clusters, we find that this is comparable to the total gas mass associated with these sources from \citetalias{Leroy18} (Table \ref{stellar}). This suggests many of these sources are still in the process of forming, and supports the assumption of a ZAMS population for the overall NGC 253 starburst, compared to the older burst age of 5 Myr adopted by \cite{Emig20} for the central starburst in NGC 4945. 

However, among the NGC 253 sources there are some indications that a range of evolutionary stages are present. We find that not all of the embedded clusters identified by \citetalias{Leroy18} have ionized gas emission coincident with the dust continuum. In particular, the new positions assigned to sources 4, 10, 11, and 12 based on the location of their H40$\alpha$ emission all have an offset (2-4 pc) from the peaks of nearby dust continuum emission identified by \citetalias{Leroy18}. Source 10NE is also significantly offset ($\sim$7 pc) from the dust continuum peaks near sources 10 and 13. These offsets are comparable to or larger than the cluster sizes, which \citetalias{Leroy18} measure to have FWHM diameters of 2.5-4.5 pc. This could indicate that these clusters are less embedded and so are more evolved than sources like 5, 8, 13 and 14 that have nearly coincident dust and free-free emission. However, the visibility of this spatial offset is a function of the orientation of the source with respect to the observer. For example, although there appears to be little offset between ionized and molecular gas in source 5, this is the only embedded cluster that is visible in near-IR HST observations \citep{Watson96,Kornei09}. This could indicate that it is one of the least embedded clusters (either because it has cleared most of its natal gas or because it is located on the front side of the molecular cloud), or alternatively it could be the only cluster visible in the near-IR because of a hole in the overall extinction screen toward the center of NGC 253.

The spatial offsets between ionized and molecular gas could also be a sign that the star formation in an individual giant molecular cloud is not entirely contemporaneous, and that the clouds may contain clusters or sub-clusters of slightly different ages. This is seen in the Milky Way center, where at high resolutions, the Sgr B2 protocluster breaks up into two separate concentrations offset by a few parsecs, with the northern (N) concentration less evolved than the main (M) concentration: it has less radio continuum emission at lower frequencies and more dust continuum emission at higher frequencies \citep{Schmiedeke16}. Even earlier stages of star formation are also seen to be distributed throughout the southernmost part of the parent molecular cloud \citep{Ginsburg18a}.

If we assign relative ages to the embedded clusters based on fraction of mass in gas versus stars, the youngest sources would be 1, 2, 3, 8, and 14, which have gas masses from \citetalias{Leroy18} that are 2-9 times larger than stellar masses we infer here (Table \ref{stellar}). For sources 1, 2, and 3 in particular, the stellar masses are more than 100 times smaller than the cluster virial mass. This is very similar to what is seen in Sgr B2, where the estimated stellar mass of the protocluster \citep[20,000-45,000 M$_\odot$;][]{Schmiedeke16,Ginsburg18a} is also  $\sim$2 orders of magnitude less than the gas mass of the whole cloud. The oldest sources would be 6, 9, 11, and 12, which all have stellar masses 4-20 times larger than the gas masses. This is consistent with the analysis of \cite{Rico-Villas20}, who find that sources 5, 6, 7, 9, 10 and 12 are more evolved than the embedded clusters 1, 2, 3, 4, 8, 13 and 14. 

We have also found that several of the embedded clusters appear to have a significant flux contribution from nonthermal radio emission, particularly sources 6, 11 and 12. This emission is most likely due to synchrotron radiation from compact supernova remnants \citep{Ulvestad97}. This could result from a chance superposition with older regions of star formation, or it could be an indication that these clusters are sufficiently evolved to have experienced several supernovae. Given that these three candidate protoclusters also have some of the lowest gas masses, we believe it is likely that these sources (as well as source 10NE, which had no counterpart in \citetalias{Leroy18} due to the lack of corresponding HCN 4$-$3 emission) represent some of the most evolved clusters associated with the NGC 253 nuclear starburst. 

\section{Conclusions}
 
We report on new, high-resolution observations of the 3 mm continuum as well as H40$\alpha$, and He40$\alpha$ line emission toward a population of embedded super star clusters in the central 200 pc of NGC 253. Below, we summarize our main findings.

\begin{itemize}
\item We measure electron temperatures for the ionized gas associated with a subset of four of the embedded clusters (4, 5, 9, and 14), and find that these range from $T_e=$ 7000-10,000 K. 
\item For three of the embedded clusters (5, 9, and 14) we also measure an average mass-weighted singly-ionized helium abundance of $\langle Y^+\rangle =0.25\pm0.6$. This is comparable to ionized helium abundances observed toward the center of the Milky Way. 
\item Using the H40$\alpha$ recombination line fluxes and the derived electron temperatures, we present revised estimates for the ionizing fluxes $Q_c$ of the embedded clusters. From these measurements we infer a slightly lower value for the total stellar mass of the clusters, which we attribute to the contribution of synchrotron emission to the radio continuum previously used to determine these values. However,  based on the increased electron temperatures we measure compared to those in \cite{Bendo15}, we now estimate that at least  50\% of the nuclear starburst in NGC 253 originates in a clustered mode of star formation. 
\item We identify 7 sources that have exceptionally broad linewidth H40$\alpha$ emission ($\Delta v_{FWHM}\sim 100-200$ \kms). These linewidths are $\sim$50 \kms\, broader than the molecular gas profiles of these sources measured from HCN 4$-3$. We suggest that these embedded clusters contribute to driving the ionized component of the multiphase outflow originating in the nucleus of NGC 253.

\end{itemize}

\section{Acknowledgements}
This paper makes use of the following ALMA data: ADS/JAO.ALMA\#2017.1.00895.S and ADS/JAO.ALMA\#2011.0.00172.S. ALMA is a partnership of ESO (representing its member states), NSF (USA) and NINS (Japan), together with NRC (Canada), MOST and ASIAA (Taiwan), and KASI (Republic of Korea), in cooperation with the Republic of Chile. The Joint ALMA Observatory is operated by ESO, AUI/NRAO and NAOJ. EACM gratefully acknowledges support by the National Science Foundation under grant No. AST-1813765. AG gratefully acknowledges support from the National Science Foundation under grant No. 2008101. RCL gratefully acknowledges support from the NSF through Student Observing Support Program (SOSP) award 7-011 from the NRAO. K.L.E. acknowledges financial support from the Netherlands Organization  for Scientific Research through TOP grant 614.001.351.

\bibliographystyle{hapj}
\bibliography{253}

\begin{thebibliography}{}
\expandafter\ifx\csname natexlab\endcsname\relax\def\natexlab#1{#1}\fi

\bibitem[{{Anantharamaiah} \& {Goss}(1996)}]{AG96}
{Anantharamaiah}, K.~R., \& {Goss}, W.~M. 1996, \apjl, 466, L13

\bibitem[{{B{\'a}ez-Rubio} {et~al.}(2018){B{\'a}ez-Rubio},
  {Mart{\'{\i}}n-Pintado}, {Rico-Villas}, \& {Jim{\'e}nez-Serra}}]{BaezRubio18}
{B{\'a}ez-Rubio}, A., {Mart{\'{\i}}n-Pintado}, J., {Rico-Villas}, F., \&
  {Jim{\'e}nez-Serra}, I. 2018, ArXiv e-prints, arXiv:1810.07213

\bibitem[{{Barnes} {et~al.}(2017){Barnes}, {Longmore}, {Battersby}, {Bally},
  {Kruijssen}, {Henshaw}, \& {Walker}}]{Barnes17}
{Barnes}, A.~T., {Longmore}, S.~N., {Battersby}, C., {et~al.} 2017, \mnras,
  469, 2263

\bibitem[{{Bauer} {et~al.}(2007){Bauer}, {Pietsch}, {Trinchieri},
  {Breitschwerdt}, {Ehle}, \& {Read}}]{Bauer07}
{Bauer}, M., {Pietsch}, W., {Trinchieri}, G., {et~al.} 2007, \aap, 467, 979

\bibitem[{{Bendo} {et~al.}(2015){Bendo}, {Beswick}, {D'Cruze}, {Dickinson},
  {Fuller}, \& {Muxlow}}]{Bendo15}
{Bendo}, G.~J., {Beswick}, R.~J., {D'Cruze}, M.~J., {et~al.} 2015, \mnras, 450,
  L80

\bibitem[{{Binney} {et~al.}(1991){Binney}, {Gerhard}, {Stark}, {Bally}, \&
  {Uchida}}]{Binney91}
{Binney}, J., {Gerhard}, O.~E., {Stark}, A.~A., {Bally}, J., \& {Uchida}, K.~I.
  1991, \mnras, 252, 210

\bibitem[{{Bolatto} {et~al.}(2013){Bolatto}, {Warren}, {Leroy}, {Walter},
  {Veilleux}, {Ostriker}, {Ott}, {Zwaan}, {Fisher}, {Weiss}, {Rosolowsky}, \&
  {Hodge}}]{Bolatto13}
{Bolatto}, A.~D., {Warren}, S.~R., {Leroy}, A.~K., {et~al.} 2013, \nat, 499,
  450

\bibitem[{{Brunthaler} {et~al.}(2009){Brunthaler}, {Castangia}, {Tarchi},
  {Henkel}, {Reid}, {Falcke}, \& {Menten}}]{Brunthaler09}
{Brunthaler}, A., {Castangia}, P., {Tarchi}, A., {et~al.} 2009, \aap, 497, 103

\bibitem[{{Christopher} {et~al.}(2005){Christopher}, {Scoville}, {Stolovy}, \&
  {Yun}}]{Chris05}
{Christopher}, M.~H., {Scoville}, N.~Z., {Stolovy}, S.~R., \& {Yun}, M.~S.
  2005, \apj, 622, 346

\bibitem[{{Churchwell} {et~al.}(1974){Churchwell}, {Mezger}, \&
  {Huchtmeier}}]{Churchwell74}
{Churchwell}, E., {Mezger}, P.~G., \& {Huchtmeier}, W. 1974, \aap, 32, 283

\bibitem[{{Clark} {et~al.}(2005){Clark}, {Negueruela}, {Crowther}, \&
  {Goodwin}}]{Clark05}
{Clark}, J.~S., {Negueruela}, I., {Crowther}, P.~A., \& {Goodwin}, S.~P. 2005,
  \aap, 434, 949

\bibitem[{{Comer{\'o}n} {et~al.}(2010){Comer{\'o}n}, {Knapen}, {Beckman},
  {Laurikainen}, {Salo}, {Mart{\'\i}nez-Valpuesta}, \& {Buta}}]{Comeron10}
{Comer{\'o}n}, S., {Knapen}, J.~H., {Beckman}, J.~E., {et~al.} 2010, \mnras,
  402, 2462

\bibitem[{{Condon}(1992)}]{Condon92}
{Condon}, J.~J. 1992, \araa, 30, 575

\bibitem[{{Dahmen} {et~al.}(1998){Dahmen}, {Huttemeister}, {Wilson}, \&
  {Mauersberger}}]{Dahmen98}
{Dahmen}, G., {Huttemeister}, S., {Wilson}, T.~L., \& {Mauersberger}, R. 1998,
  \aap, 331, 959

\bibitem[{{Davidge}(2016)}]{Davidge16}
{Davidge}, T.~J. 2016, \apj, 818, 142

\bibitem[{{de Pree} {et~al.}(1996){de Pree}, {Gaume}, {Goss}, \&
  {Claussen}}]{dePree96}
{de Pree}, C.~G., {Gaume}, R.~A., {Goss}, W.~M., \& {Claussen}, M.~J. 1996,
  \apj, 464, 788

\bibitem[{{Draine}(2011)}]{Draine}
{Draine}, B.~T. 2011, {Physics of the Interstellar and Intergalactic Medium}

\bibitem[{{Emig} {et~al.}(2020){Emig}, {Bolatto}, {Leroy}, {Mills}, {Jimenez
  Donaire}, {Tielens}, {Ginsburg}, {Gorski}, {Krieger}, {Levy}, {Meier}, {Ott},
  {Rosolowsky}, {Thompson}, \& {Veilleux}}]{Emig20}
{Emig}, K.~L., {Bolatto}, A.~D., {Leroy}, A.~K., {et~al.} 2020, arXiv e-prints,
  arXiv:2009.05154

\bibitem[{{Engelbracht} {et~al.}(1998){Engelbracht}, {Rieke}, {Rieke}, {Kelly},
  \& {Achtermann}}]{Engelbracht98}
{Engelbracht}, C.~W., {Rieke}, M.~J., {Rieke}, G.~H., {Kelly}, D.~M., \&
  {Achtermann}, J.~M. 1998, \apj, 505, 639

\bibitem[{{Fern{\'a}ndez-Ontiveros} {et~al.}(2009){Fern{\'a}ndez-Ontiveros},
  {Prieto}, \& {Acosta-Pulido}}]{FO09}
{Fern{\'a}ndez-Ontiveros}, J.~A., {Prieto}, M.~A., \& {Acosta-Pulido}, J.~A.
  2009, \mnras, 392, L16

\bibitem[{{Ghez} {et~al.}(2008)}]{Ghez08}
{Ghez}, A.~M., {et~al.} 2008, \apj, 689, 1044

\bibitem[{{Ginsburg} {et~al.}(2015){Ginsburg}, {Bally}, {Battersby},
  {Youngblood}, {Darling}, {Rosolowsky}, {Arce}, \& {Lebr{\'o}n
  Santos}}]{Ginsburg15}
{Ginsburg}, A., {Bally}, J., {Battersby}, C., {et~al.} 2015, \aap, 573, A106

\bibitem[{{Ginsburg} \& {Kruijssen}(2018)}]{Ginsburg18}
{Ginsburg}, A., \& {Kruijssen}, J.~M.~D. 2018, \apjl, 864, L17

\bibitem[{{Ginsburg} \& {Mirocha}(2011)}]{Ginsburg11}
{Ginsburg}, A., \& {Mirocha}, J. 2011, {PySpecKit: Python Spectroscopic
  Toolkit}, ascl:1109.001

\bibitem[{{Ginsburg} {et~al.}(2018){Ginsburg}, {Bally}, {Barnes}, {Bastian},
  {Battersby}, {Beuther}, {Brogan}, {Contreras}, {Corby}, {Darling}, {De Pree},
  {Galv{\'a}n-Madrid}, {Garay}, {Henshaw}, {Hunter}, {Kruijssen}, {Longmore},
  {Lu}, {Meng}, {Mills}, {Ott}, {Pineda}, {S{\'a}nchez-Monge}, {Schilke},
  {Schmiedeke}, {Walker}, \& {Wilner}}]{Ginsburg18a}
{Ginsburg}, A., {Bally}, J., {Barnes}, A., {et~al.} 2018, \apj, 853, 171

\bibitem[{{Gordon} \& {Sorochenko}(2002)}]{GS02}
{Gordon}, M.~A., \& {Sorochenko}, R.~L. 2002, {Radio Recombination Lines. Their
  Physics and Astronomical Applications}, Vol. 282,
  doi:10.1007/978-0-387-09604-9

\bibitem[{{Gorski} {et~al.}(2017){Gorski}, {Ott}, {Rand}, {Meier}, {Momjian},
  \& {Schinnerer}}]{Gorski17}
{Gorski}, M., {Ott}, J., {Rand}, R., {et~al.} 2017, \apj, 842, 124

\bibitem[{{Gorski} {et~al.}(2019){Gorski}, {Ott}, {Rand}, {Meier}, {Momjian},
  {Schinnerer}, \& {Ellingsen}}]{Gorski19}
{Gorski}, M.~D., {Ott}, J., {Rand}, R., {et~al.} 2019, \mnras, 483, 5434

\bibitem[{{Goss} {et~al.}(1985){Goss}, {Schwarz}, {van Gorkom}, \&
  {Ekers}}]{Goss85}
{Goss}, W.~M., {Schwarz}, U.~J., {van Gorkom}, J.~H., \& {Ekers}, R.~D. 1985,
  \mnras, 215, 69P

\bibitem[{{Henkel} {et~al.}(2004){Henkel}, {Tarchi}, {Menten}, \&
  {Peck}}]{Henkel04}
{Henkel}, C., {Tarchi}, A., {Menten}, K.~M., \& {Peck}, A.~B. 2004, \aap, 414,
  117

\bibitem[{{Henshaw} {et~al.}(2016){Henshaw}, {Longmore}, {Kruijssen}, {Davies},
  {Bally}, {Barnes}, {Battersby}, {Burton}, {Cunningham}, {Dale}, {Ginsburg},
  {Immer}, {Jones}, {Kendrew}, {Mills}, {Molinari}, {Moore}, {Ott}, {Pillai},
  {Rathborne}, {Schilke}, {Schmiedeke}, {Testi}, {Walker}, {Walsh}, \&
  {Zhang}}]{Henshaw16}
{Henshaw}, J.~D., {Longmore}, S.~N., {Kruijssen}, J.~M.~D., {et~al.} 2016,
  \mnras, 457, 2675

\bibitem[{{Hensley} {et~al.}(2015){Hensley}, {Murphy}, \&
  {Staguhn}}]{Hensley15}
{Hensley}, B., {Murphy}, E., \& {Staguhn}, J. 2015, \mnras, 449, 809

\bibitem[{{Izumi} {et~al.}(2016){Izumi}, {Nakanishi}, {Imanishi}, \&
  {Kohno}}]{Izumi16}
{Izumi}, T., {Nakanishi}, K., {Imanishi}, M., \& {Kohno}, K. 2016, \mnras, 459,
  3629

\bibitem[{{Kauffmann} {et~al.}(2017){Kauffmann}, {Goldsmith}, {Melnick},
  {Tolls}, {Guzman}, \& {Menten}}]{Kauffmann17}
{Kauffmann}, J., {Goldsmith}, P.~F., {Melnick}, G., {et~al.} 2017, \aap, 605,
  L5

\bibitem[{{Kornei} \& {McCrady}(2009)}]{Kornei09}
{Kornei}, K.~A., \& {McCrady}, N. 2009, \apj, 697, 1180

\bibitem[{{Krieger} {et~al.}(2019){Krieger}, {Bolatto}, {Walter}, {Leroy},
  {Zschaechner}, {Meier}, {Ott}, {Weiss}, {Mills}, {Levy}, {Veilleux}, \&
  {Gorski}}]{Krieger19}
{Krieger}, N., {Bolatto}, A.~D., {Walter}, F., {et~al.} 2019, \apj, 881, 43

\bibitem[{{Kruijssen} {et~al.}(2015){Kruijssen}, {Dale}, \&
  {Longmore}}]{Kruijssen15}
{Kruijssen}, J.~M.~D., {Dale}, J.~E., \& {Longmore}, S.~N. 2015, \mnras, 447,
  1059

\bibitem[{{Krumholz} \& {Kruijssen}(2015)}]{KK15}
{Krumholz}, M.~R., \& {Kruijssen}, J.~M.~D. 2015, \mnras, 453, 739

\bibitem[{{Lang} {et~al.}(2001){Lang}, {Goss}, \& {Morris}}]{Lang01}
{Lang}, C.~C., {Goss}, W.~M., \& {Morris}, M. 2001, \aj, 121, 2681

\bibitem[{{Lang} {et~al.}(1997){Lang}, {Goss}, \& {Wood}}]{Lang97}
{Lang}, C.~C., {Goss}, W.~M., \& {Wood}, O.~S. 1997, \apj, 474, 275

\bibitem[{{Launhardt} {et~al.}(2002){Launhardt}, {Zylka}, \&
  {Mezger}}]{Launhardt02}
{Launhardt}, R., {Zylka}, R., \& {Mezger}, P.~G. 2002, \aap, 384, 112

\bibitem[{{Leitherer} {et~al.}(1999){Leitherer}, {Schaerer}, {Goldader},
  {Delgado}, {Robert}, {Kune}, {de Mello}, {Devost}, \&
  {Heckman}}]{Leitherer99}
{Leitherer}, C., {Schaerer}, D., {Goldader}, J.~D., {et~al.} 1999, \apjs, 123,
  3

\bibitem[{{Leroy} {et~al.}(2015){Leroy}, {Bolatto}, {Ostriker}, {Rosolowsky},
  {Walter}, {Warren}, {Donovan Meyer}, {Hodge}, {Meier}, {Ott}, {Sandstrom},
  {Schruba}, {Veilleux}, \& {Zwaan}}]{Leroy15}
{Leroy}, A.~K., {Bolatto}, A.~D., {Ostriker}, E.~C., {et~al.} 2015, \apj, 801,
  25

\bibitem[{{Leroy} {et~al.}(2018){Leroy}, {Bolatto}, {Ostriker}, {Walter},
  {Gorski}, {Ginsburg}, {Krieger}, {Meier}, {Mills}, {Ott}, {Rosolowsky},
  {Thompson}, {Veilleux}, \& {Zschaechner}}]{Leroy18}
---. 2018, ArXiv e-prints, arXiv:1804.02083

\bibitem[{{Levesque} \& {Leitherer}(2013)}]{Levesque13}
{Levesque}, E.~M., \& {Leitherer}, C. 2013, \apj, 779, 170

\bibitem[{{Levesque} {et~al.}(2012){Levesque}, {Leitherer}, {Ekstrom},
  {Meynet}, \& {Schaerer}}]{Levesque12}
{Levesque}, E.~M., {Leitherer}, C., {Ekstrom}, S., {Meynet}, G., \& {Schaerer},
  D. 2012, \apj, 751, 67

\bibitem[{{Levy} {et~al.}(2021){Levy}, {Bolatto}, {Leroy}, {Emig}, {Gorski},
  {Krieger}, {Lenki{\'c}}, {Meier}, {Mills}, {Ott}, {Rosolowsky}, {Tarantino},
  {Veilleux}, {Walter}, {Wei{\ss}}, \& {Zwaan}}]{Levy21}
{Levy}, R.~C., {Bolatto}, A.~D., {Leroy}, A.~K., {et~al.} 2021, \apj, 912, 4

\bibitem[{{Longmore} {et~al.}(2013){Longmore}, {Bally}, {Testi}, {Purcell},
  {Walsh}, {Bressert}, {Pestalozzi}, {Molinari}, {Ott}, {Cortese}, {Battersby},
  {Murray}, {Lee}, {Kruijssen}, {Schisano}, \& {Elia}}]{Longmore13}
{Longmore}, S.~N., {Bally}, J., {Testi}, L., {et~al.} 2013, \mnras, 429, 987

\bibitem[{{Marconi} {et~al.}(2000){Marconi}, {Oliva}, {van der Werf},
  {Maiolino}, {Schreier}, {Macchetto}, \& {Moorwood}}]{Marconi00}
{Marconi}, A., {Oliva}, E., {van der Werf}, P.~P., {et~al.} 2000, \aap, 357, 24

\bibitem[{{Mauersberger} {et~al.}(1996){Mauersberger}, {Henkel}, {Wielebinski},
  {Wiklind}, \& {Reuter}}]{Mauersberger96}
{Mauersberger}, R., {Henkel}, C., {Wielebinski}, R., {Wiklind}, T., \&
  {Reuter}, H.-P. 1996, \aap, 305, 421

\bibitem[{{Meier} {et~al.}(2002){Meier}, {Turner}, \& {Beck}}]{Meier02}
{Meier}, D.~S., {Turner}, J.~L., \& {Beck}, S.~C. 2002, \aj, 124, 877

\bibitem[{{Meier} {et~al.}(2015){Meier}, {Walter}, {Bolatto}, {Leroy}, {Ott},
  {Rosolowsky}, {Veilleux}, {Warren}, {Wei{\ss}}, {Zwaan}, \&
  {Zschaechner}}]{Meier15}
{Meier}, D.~S., {Walter}, F., {Bolatto}, A.~D., {et~al.} 2015, \apj, 801, 63

\bibitem[{{M{\'e}ndez-Delgado} {et~al.}(2020){M{\'e}ndez-Delgado}, {Esteban},
  {Garc{\'\i}a-Rojas}, {Arellano-C{\'o}rdova}, \& {Valerdi}}]{MendezDelgado20}
{M{\'e}ndez-Delgado}, J.~E., {Esteban}, C., {Garc{\'\i}a-Rojas}, J.,
  {Arellano-C{\'o}rdova}, K.~Z., \& {Valerdi}, M. 2020, \mnras, 496, 2726

\bibitem[{{Mengel} {et~al.}(2002){Mengel}, {Lehnert}, {Thatte}, \&
  {Genzel}}]{Mengel02}
{Mengel}, S., {Lehnert}, M.~D., {Thatte}, N., \& {Genzel}, R. 2002, \aap, 383,
  137

\bibitem[{{Mezger} \& {Smith}(1976)}]{Mezger76}
{Mezger}, P.~G., \& {Smith}, L.~F. 1976, \aap, 47, 143

\bibitem[{{Mezger} {et~al.}(1974){Mezger}, {Smith}, \& {Churchwell}}]{Mezger74}
{Mezger}, P.~G., {Smith}, L.~F., \& {Churchwell}, E. 1974, \aap, 32, 269

\bibitem[{{Mills} \& {Battersby}(2017)}]{Mills17a}
{Mills}, E.~A.~C., \& {Battersby}, C. 2017, \apj, 835, 76

\bibitem[{{Mills} {et~al.}(2018){Mills}, {Ginsburg}, {Clements}, {Schilke},
  {S{\'a}nchez-Monge}, {Menten}, {Butterfield}, {Goddi}, {Schmiedeke}, \& {De
  Pree}}]{Mills18c}
{Mills}, E.~A.~C., {Ginsburg}, A., {Clements}, A.~R., {et~al.} 2018, \apjl,
  869, L14

\bibitem[{{Mohan} {et~al.}(2005){Mohan}, {Goss}, \& {Anantharamaiah}}]{Mohan05}
{Mohan}, N.~R., {Goss}, W.~M., \& {Anantharamaiah}, K.~R. 2005, \aap, 432, 1

\bibitem[{{Molinari} {et~al.}(2011){Molinari}, {Bally}, {Noriega-Crespo},
  {Compi{\`e}gne}, {Bernard}, {Paradis}, {Martin}, {Testi}, {Barlow}, {Moore},
  {Plume}, {Swinyard}, {Zavagno}, {Calzoletti}, {Di Giorgio}, {Elia},
  {Faustini}, {Natoli}, {Pestalozzi}, {Pezzuto}, {Piacentini}, {Polenta},
  {Polychroni}, {Schisano}, {Traficante}, {Veneziani}, {Battersby}, {Burton},
  {Carey}, {Fukui}, {Li}, {Lord}, {Morgan}, {Motte}, {Schuller},
  {Stringfellow}, {Tan}, {Thompson}, {Ward-Thompson}, {White}, \&
  {Umana}}]{Molinari11}
{Molinari}, S., {Bally}, J., {Noriega-Crespo}, A., {et~al.} 2011, \apjl, 735,
  L33

\bibitem[{{M{\"u}ller-S{\'a}nchez} {et~al.}(2010){M{\"u}ller-S{\'a}nchez},
  {Gonz{\'a}lez-Mart{\'{\i}}n}, {Fern{\'a}ndez-Ontiveros}, {Acosta-Pulido}, \&
  {Prieto}}]{MullerSanchez10}
{M{\"u}ller-S{\'a}nchez}, F., {Gonz{\'a}lez-Mart{\'{\i}}n}, O.,
  {Fern{\'a}ndez-Ontiveros}, J.~A., {Acosta-Pulido}, J.~A., \& {Prieto}, M.~A.
  2010, \apj, 716, 1166

\bibitem[{{Ott} {et~al.}(2005){Ott}, {Weiss}, {Henkel}, \& {Walter}}]{Ott05}
{Ott}, J., {Weiss}, A., {Henkel}, C., \& {Walter}, F. 2005, \apj, 629, 767

\bibitem[{{Pietsch} {et~al.}(2000){Pietsch}, {Vogler}, {Klein}, \&
  {Zinnecker}}]{Pietsch00}
{Pietsch}, W., {Vogler}, A., {Klein}, U., \& {Zinnecker}, H. 2000, \aap, 360,
  24

\bibitem[{{Protheroe} {et~al.}(2008){Protheroe}, {Ott}, {Ekers}, {Jones}, \&
  {Crocker}}]{Protheroe08}
{Protheroe}, R.~J., {Ott}, J., {Ekers}, R.~D., {Jones}, D.~I., \& {Crocker},
  R.~M. 2008, \mnras, 390, 683

\bibitem[{{Prozesky} \& {Smits}(2020)}]{Prozesky20}
{Prozesky}, A., \& {Smits}, D.~P. 2020, \mnras, 491, 2536

\bibitem[{{Quireza} {et~al.}(2006){Quireza}, {Rood}, {Bania}, {Balser}, \&
  {Maciel}}]{Quireza06}
{Quireza}, C., {Rood}, R.~T., {Bania}, T.~M., {Balser}, D.~S., \& {Maciel},
  W.~J. 2006, \apj, 653, 1226

\bibitem[{{Rekola} {et~al.}(2005){Rekola}, {Richer}, {McCall}, {Valtonen},
  {Kotilainen}, \& {Flynn}}]{Rekola05}
{Rekola}, R., {Richer}, M.~G., {McCall}, M.~L., {et~al.} 2005, \mnras, 361, 330

\bibitem[{{Rico-Villas} {et~al.}(2020){Rico-Villas}, {Mart{\'\i}n-Pintado},
  {Gonz{\'a}lez-Alfonso}, {Mart{\'\i}n}, \& {Rivilla}}]{Rico-Villas20}
{Rico-Villas}, F., {Mart{\'\i}n-Pintado}, J., {Gonz{\'a}lez-Alfonso}, E.,
  {Mart{\'\i}n}, S., \& {Rivilla}, V.~M. 2020, \mnras, 491, 4573

\bibitem[{{Rodr{\'{\i}}guez-Rico} {et~al.}(2006){Rodr{\'{\i}}guez-Rico},
  {Goss}, {Zhao}, {G{\'o}mez}, \& {Anantharamaiah}}]{RodriguezRico06}
{Rodr{\'{\i}}guez-Rico}, C.~A., {Goss}, W.~M., {Zhao}, J.-H., {G{\'o}mez}, Y.,
  \& {Anantharamaiah}, K.~R. 2006, \apj, 644, 914

\bibitem[{{Rubin}(1968)}]{Rubin68}
{Rubin}, R.~H. 1968, \apj, 154, 391

\bibitem[{{Sakamoto} {et~al.}(2011){Sakamoto}, {Mao}, {Matsushita}, {Peck},
  {Sawada}, \& {Wiedner}}]{Sakamoto11}
{Sakamoto}, K., {Mao}, R.-Q., {Matsushita}, S., {et~al.} 2011, \apj, 735, 19

\bibitem[{{S{\'a}nchez-Monge} {et~al.}(2017){S{\'a}nchez-Monge}, {Schilke},
  {Schmiedeke}, {Ginsburg}, {Cesaroni}, {Lis}, {Qin}, {M{\"u}ller}, {Bergin},
  {Comito}, \& {M{\"o}ller}}]{SM17}
{S{\'a}nchez-Monge}, {\'A}., {Schilke}, P., {Schmiedeke}, A., {et~al.} 2017,
  \aap, 604, A6

\bibitem[{Schmiedeke {et~al.}(2016)Schmiedeke, Schilke, M{\"o}ller,
  S{\'a}nchez-Monge, Bergin, Comito, Csengeri, Lis, Molinari, Qin, \&
  Rolffs}]{Schmiedeke16}
Schmiedeke, A., Schilke, P., M{\"o}ller, T., {et~al.} 2016, \aap, 588, A143

\bibitem[{{Scoville} \& {Murchikova}(2013)}]{Scoville13}
{Scoville}, N., \& {Murchikova}, L. 2013, \apj, 779, 75

\bibitem[{{Sharp} \& {Bland-Hawthorn}(2010)}]{Sharp10}
{Sharp}, R.~G., \& {Bland-Hawthorn}, J. 2010, \apj, 711, 818

\bibitem[{{Sormani} {et~al.}(2015){Sormani}, {Binney}, \&
  {Magorrian}}]{Sormani15}
{Sormani}, M.~C., {Binney}, J., \& {Magorrian}, J. 2015, \mnras, 449, 2421

\bibitem[{{Sormani} \& {Li}(2020)}]{Sormani20}
{Sormani}, M.~C., \& {Li}, Z. 2020, arXiv e-prints, arXiv:2002.10559

\bibitem[{{Sormani} {et~al.}(2020){Sormani}, {Magorrian}, {Nogueras-Lara},
  {Neumayer}, {Sch{\"o}nrich}, {Klessen}, \&
  {Mastrobuono-Battisti}}]{Sormani20b}
{Sormani}, M.~C., {Magorrian}, J., {Nogueras-Lara}, F., {et~al.} 2020, \mnras,
  499, 7

\bibitem[{{Spoon} {et~al.}(2000){Spoon}, {Koornneef}, {Moorwood}, {Lutz}, \&
  {Tielens}}]{Spoon00}
{Spoon}, H.~W.~W., {Koornneef}, J., {Moorwood}, A.~F.~M., {Lutz}, D., \&
  {Tielens}, A.~G.~G.~M. 2000, \aap, 357, 898

\bibitem[{{Storey} \& {Hummer}(1995)}]{SH95}
{Storey}, P.~J., \& {Hummer}, D.~G. 1995, \mnras, 272, 41

\bibitem[{{Strickland} {et~al.}(2002){Strickland}, {Heckman}, {Weaver},
  {Hoopes}, \& {Dahlem}}]{Strickland02}
{Strickland}, D.~K., {Heckman}, T.~M., {Weaver}, K.~A., {Hoopes}, C.~G., \&
  {Dahlem}, M. 2002, \apj, 568, 689

\bibitem[{{Sugai} {et~al.}(2003){Sugai}, {Davies}, \& {Ward}}]{Sugai03}
{Sugai}, H., {Davies}, R.~I., \& {Ward}, M.~J. 2003, \apjl, 584, L9

\bibitem[{{Torrey} {et~al.}(2017){Torrey}, {Hopkins}, {Faucher-Gigu{\`e}re},
  {Vogelsberger}, {Quataert}, {Kere{\v{s}}}, \& {Murray}}]{Torrey17}
{Torrey}, P., {Hopkins}, P.~F., {Faucher-Gigu{\`e}re}, C.-A., {et~al.} 2017,
  \mnras, 467, 2301

\bibitem[{{Turner} {et~al.}(2000){Turner}, {Beck}, \& {Ho}}]{Turner00}
{Turner}, J.~L., {Beck}, S.~C., \& {Ho}, P. T.~P. 2000, \apjl, 532, L109

\bibitem[{{Turner} \& {Ho}(1985)}]{TurnerHo85}
{Turner}, J.~L., \& {Ho}, P.~T.~P. 1985, \apjl, 299, L77

\bibitem[{{Turner} {et~al.}(1998){Turner}, {Ho}, \& {Beck}}]{Turner98}
{Turner}, J.~L., {Ho}, P. T.~P., \& {Beck}, S.~C. 1998, \aj, 116, 1212

\bibitem[{{Ulvestad} \& {Antonucci}(1997)}]{Ulvestad97}
{Ulvestad}, J.~S., \& {Antonucci}, R.~R.~J. 1997, \apj, 488, 621

\bibitem[{van~der Walt {et~al.}(2014)van~der Walt, Schonberger, Nunez-Iglesias,
  Boulogne, Warner, Yager, Gouillart, \& Yu}]{scikit14}
van~der Walt, S., Schonberger, J.~L., Nunez-Iglesias, J., {et~al.} 2014, PeerJ,
  2, e453

\bibitem[{{Veilleux} {et~al.}(2020){Veilleux}, {Maiolino}, {Bolatto}, \&
  {Aalto}}]{Veilleux20}
{Veilleux}, S., {Maiolino}, R., {Bolatto}, A.~D., \& {Aalto}, S. 2020, \aapr,
  28, 2

\bibitem[{{Walter} {et~al.}(2017){Walter}, {Bolatto}, {Leroy}, {Veilleux},
  {Warren}, {Hodge}, {Levy}, {Meier}, {Ostriker}, {Ott}, {Rosolowsky},
  {Scoville}, {Weiss}, {Zschaechner}, \& {Zwaan}}]{Walter17}
{Walter}, F., {Bolatto}, A.~D., {Leroy}, A.~K., {et~al.} 2017, \apj, 835, 265

\bibitem[{{Watson} {et~al.}(1996){Watson}, {Gallagher}, {Holtzman}, {Hester},
  {Mould}, {Ballester}, {Burrows}, {Casertano}, {Clarke}, {Crisp}, {Evans},
  {Griffiths}, {Hoessel}, {Scowen}, {Stapelfeldt}, {Trauger}, \&
  {Westphal}}]{Watson96}
{Watson}, A.~M., {Gallagher}, J.~S., I., {Holtzman}, J.~A., {et~al.} 1996, \aj,
  112, 534

\bibitem[{{Wenger} {et~al.}(2013){Wenger}, {Bania}, {Balser}, \&
  {Anderson}}]{Wenger13}
{Wenger}, T.~V., {Bania}, T.~M., {Balser}, D.~S., \& {Anderson}, L.~D. 2013,
  \apj, 764, 34

\bibitem[{{Westmoquette} {et~al.}(2011){Westmoquette}, {Smith}, \&
  {Gallagher}}]{Westmoquette11}
{Westmoquette}, M.~S., {Smith}, L.~J., \& {Gallagher}, III, J.~S. 2011, \mnras,
  414, 3719

\bibitem[{{Wynn-Williams} {et~al.}(1979){Wynn-Williams}, {Becklin}, {Matthews},
  \& {Neugebauer}}]{WynnWilliams79}
{Wynn-Williams}, C.~G., {Becklin}, E.~E., {Matthews}, K., \& {Neugebauer}, G.
  1979, \mnras, 189, 163

\bibitem[{{Zhao} \& {Wright}(2011)}]{Zhao11}
{Zhao}, J.-H., \& {Wright}, M.~C.~H. 2011, \apj, 742, 50

\bibitem[{{Zschaechner} {et~al.}(2018){Zschaechner}, {Bolatto}, {Walter},
  {Leroy}, {Herrera}, {Krieger}, {Kruijssen}, {Meier}, {Mills}, {Ott},
  {Veilleux}, \& {Weiss}}]{Zschaechner18}
{Zschaechner}, L.~K., {Bolatto}, A.~D., {Walter}, F., {et~al.} 2018, \apj, 867,
  111

\end{thebibliography}

\end{document}